# Giant near-field nonlinear electrophotonic effects in an angstrom-scale plasmonic junction

Shota Takahashi[1], Atsunori Sakurai[1,2*], Tatsuto Mochizuki[1,2], and Toshiki Sugimoto[1,2*]

[1] Institute for Molecular Science, National Institutes of Natural Sciences; Okazaki, Aichi 444-8585, Japan.

[2] Graduate Institute for Advanced Studies, SOKENDAI; Okazaki, Aichi 444-8585, Japan.

*Corresponding authors. Email: asakurai@ims.ac.jp; toshiki-sugimoto@ims.ac.jp

## Abstract

Plasmons facilitate a strong confinement and enhancement of near-field light, offering exciting opportunities to enhance nonlinear optical responses at the nanoscale. However, despite significant advancements, the electrically tunable range of the nonlinear optical responses at nanometer-scale plasmonic structures remains limited to a few percents per volt. Here, we transcend the limitation of the nanometer regime by expanding the concept of electrophotonics into angstrom-scale platform, enabling high-performance modulation of near-field nonlinear optical responses inaccessible in prior architectures. We demonstrate ~2000% enhancement in second-harmonic generation (SHG) within 1 V of voltage application by utilizing an angstrom-scale plasmonic gap between a metallic tip and a flat metal substrate in a scanning tunneling microscope. Extending this near-field SHG scheme to sum-frequency generation that is accompanied by large frequency upconversion, we also found that such giant electrical modulation of plasmon-enhanced nonlinear optical phenomena is effective over mid-infrared to visible broad wavelength range. Our results and concepts lay the foundation for developing near-field-based angstrom-scale nonlinear electrophotonics with significant modulation depth at low driving voltage.

An ultimate goal in photonics is to design photonic phenomena with desired functions by intentionally controlling light-matter interactions. While the photonic functionalities are fundamentally determined and limited by the intrinsic material properties such as permittivity and conductivity, much effort has been devoted to tune and exploit those functionalities. In particular, tailoring nonlinear optical effects is of paramount importance with the growing demand for diverse photonic functionalities, such as light frequency conversion[1], all-optical switching[2], optical imaging[3], and spectroscopic analysis of materials[4,5]. Since the nonlinear optical processes depend superlinearly on the incident electric field, integrating with plasmons that intensely confine and enhance light within nanometric volumes should dramatically augment near-field nonlinear optical phenomena at the nanoscale beyond the diffraction limit. This concept underpins a fascinating research field of nonlinear plasmonics[6–10], encompassing various applications such as enhanced up/down conversion of optical frequency[11–13], plasmonic sensing of local environments[14–17], and nonlinear optical imaging of nanomaterials[18–25]. Nonetheless, active control of plasmonic nonlinear optical effects is often constrained by limited tunability. Plasmonic functionalities are primarily governed by the intrinsic optical properties of constituent materials and the static geometry of the plasmonic structures (e.g. size, shape, material composition, and surrounding environment). Once these structural parameters are fully fixed, the plasmonic response becomes uniquely determined, rendering post-fabrication adjustment inherently challenging. Consequently, although considerable efforts have been devoted to control plasmonic properties by tuning the size, shape, and material content[26–31], efficient post-fabrication tuning of plasmonic nonlinear optical properties still remains an open challenge.

One potential strategy to overcome this constraint involves leveraging the metallic nature of plasmonic systems. Metal plasmonic structures can serve not only as light-enhancing media in near-field optical scheme but also as electrodes that can facilitate various electrical perturbations[32–36]. These dual optical and electrical functionalities provide an interesting platform for electrical modulation of plasmonic nonlinear optical processes, which are critical for on-chip integrated optoelectronics including tunable nanolasers[37–39] and optical modulators[40,41]. By applying the voltage across plasmonic gap structures, an electrostatic field is generated within these gaps, which can affect the plasmonic nonlinear optical properties inside the gaps. Although several studies have successfully reported the possibility of the plasmonic gap-based modulation[42–46], most have focused on relatively wide (sub-100 nm) gap structures, as fabricating and maintaining angstrom-scale narrow gaps remains a significant technical challenge. Consequently, while valuable progress has been made, the reported modulation depths have typically been limited to less than a 10% signal increase per volt[42–46]. Therefore, a practical level of electric modulation reaching 1000% signal increase has typically required as high as ~100 V of voltage application, preventing the application to realistic device.

In this study, we expand the platform of electrophotonics from the sub-100-nanometer into angstrom regime, unveiling the potential of an angstrom-scale plasmonic junction as a groundbreaking platform for electrophotonic control of near-field nonlinear optical effects. As a first experimental realization of such angstrom-scale electrophotonics, we leverage a voltage-applicable and precisely position-controllable plasmonic gold tip integrated with a scanning tunneling microscope (STM). This system achieves giant electrical modulation of plasmonic nonlinear optical responses, demonstrating a ~2000% signal enhancement with a minimal voltage sweep of 1 V. Our work represents the first experimental realization of this near-field nonlinear electrophotonic effect in second-harmonic generation (SHG) excited by femto-second near-infrared (IR) laser pulses. Moreover, by temporally and spatially superimposing the near- and mid-IR laser pulses, we extend this effect to the sum-frequency generation (SFG) process that is accompanied by large frequency upconversion from mid-IR to visible region. Showing broadband operability of the gigantic near-field nonlinear electrophotonic effects, our work shows great potential for advancing functionalities in angstrom-scale electrophotonic systems.

## Results and discussion

**Angstrom-scale platform of nonlinear plasmonics.** SHG is a second-order nonlinear optical process in which two photons with identical frequencies interact with a material and are converted into a single photon with doubled frequency of the original photons (Fig. 1a). The possibility of electric modulation of the near-field SHG signals from angstrom-scale plasmonic gap was explored in the experimental setup illustrated in Fig. 1c. For our first demonstration, we adopted a recently constructed experimental platform based on an angstrom-scale gap that consisted of an electrochemically etched Au tip[47] (Fig. 1b) and an atomically flat Au(111) substrate equipped in an ultra-high vacuum STM unit ($<1 \times 10^{-7}$ Pa)[48,49]. In this system, we can control the distance of plasmonic gap between the tip apex and the substrate at the angstrom scale by precisely regulating the tunneling current (further details in Supplementary Sections 3 and 4). Prior to the SHG experiments, the tip underwent $Ar^+$ sputtering for 3 hours to clean the tip apex[50]. This sputtering enabled reproducible nonlinear optical experiments presented below, as well as clear STM imaging.

In the gap, ~6-Å-thick self-assembled monolayer (SAM) of 4-methylbenzenthiol (MBT) was formed on the Au substrate as a model ultrathin dielectric layer with well-defined homogenous structure[51] (See Supplementary Fig. S1 for the STM image of a MBT SAM). To initiate the tip-enhanced SHG (TE-SHG) process, we exposed the gap to near-IR excitation pulses (1500 nm, 280 fs, FWHM: 19 nm) with *p*-polarization at a high repetition rate (50 MHz). In general, when the optical structure is considerably smaller than the optical wavelength, the phase-matching condition breaks down, resulting in a dipole-like radiation pattern of the output signal[18–22]. Thus, we collected TE-SHG emissions separately in forward-and backward-scattering directions (see Method section for details). As reported in our recent study[48], the TE-SHG output exhibits long-time stability and quadratic dependence on the excitation intensity provided the excitation power is maintained sufficiently low (≤0.7 mW). Therefore, to ensure minimal signal fluctuations, the excitation intensity in all TE-SHG measurements presented in the subsequent sections was set to 0.5 mW (10 pJ/pulse). Indeed, employing this intensity allowed us to obtain clear STM images even under light irradiation (Supplementary Fig. S11), ensuring that the optical damaging effect induced by laser pulse is sufficiently suppressed[48].

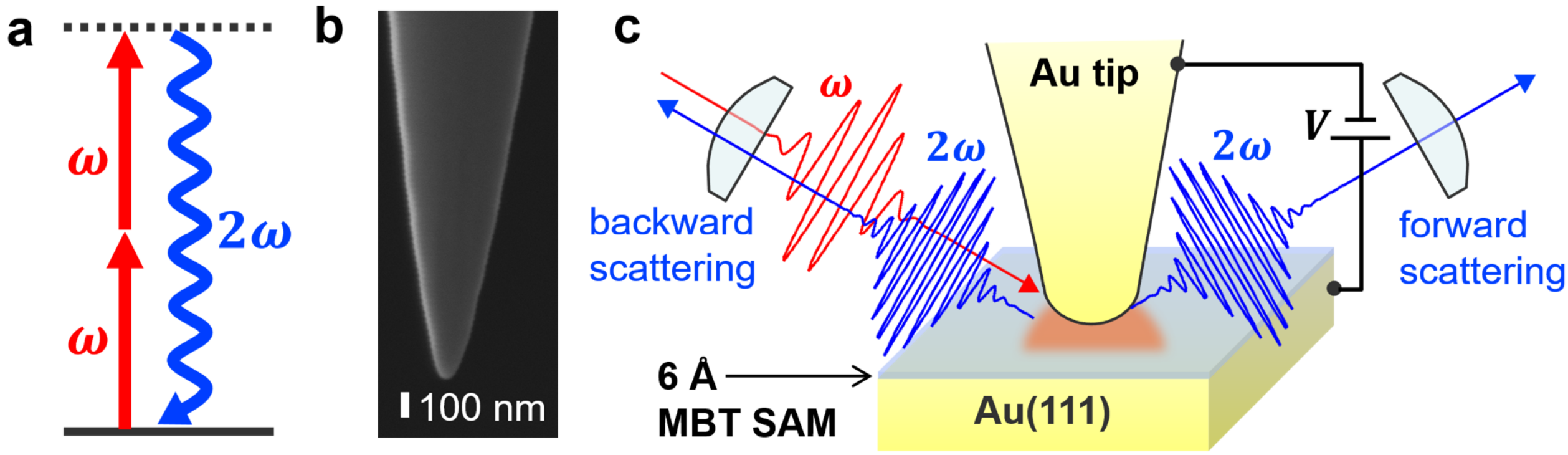

**Fig. 1 | Near-field SHG experiments in the angstrom-scale plasmonic junction of STM. a.** Energy level diagram of SHG process. **b.** Scanning electron micrograph of the Au tip used in the experiments. **c.** Schematic depiction of near-field SHG experiment conducted under room temperature and ultra-high vacuum (<1 × $10^{-7}$ Pa) condition. A Au tip and a Au substrate were mounted on an STM unit and formed an angstrom-scale gap with an applied bias voltage $V$. An ultrathin (~6 Å) dielectric layer composed of MBT SAM was formed on the Au substrate. The gap region was irradiated by femtosecond near-IR laser with a frequency of $\omega$, and near-field SHG with a frequency of $2\omega$ was detected in both forward- and backward-scattering geometries.

**Near-field SHG and its giant electric modulation.** When the substrate was retracted from the tip by ~30 nm, thereby deactivating the plasmonic excitation, no backscattered SHG signal was detected (gray curve in Fig. 2a). Owing to the phase-matching condition of far-field nonlinear optical processes, normal far-field SHG signal from the flat Au(111) substrate without plasmonic excitation was detected only in the forward-scattering direction (Supplementary Fig. S7). Although tip plasmons of the tip apex alone could be excited even in the 30-nm-retracted condition, their contribution can be safely disregarded because the backward-scattered signal emission is absent despite identical power and integration time for the measurements of forward- and backward-scattered signals (further details in Supplementary Section 5). The minimal contribution from the tip plasmon was further corroborated by our numerical simulation, revealing that the plasmonic enhancement caused by a tip alone in the absence of an angstrom-scale gap is extremely small (Supplementary Section 15).

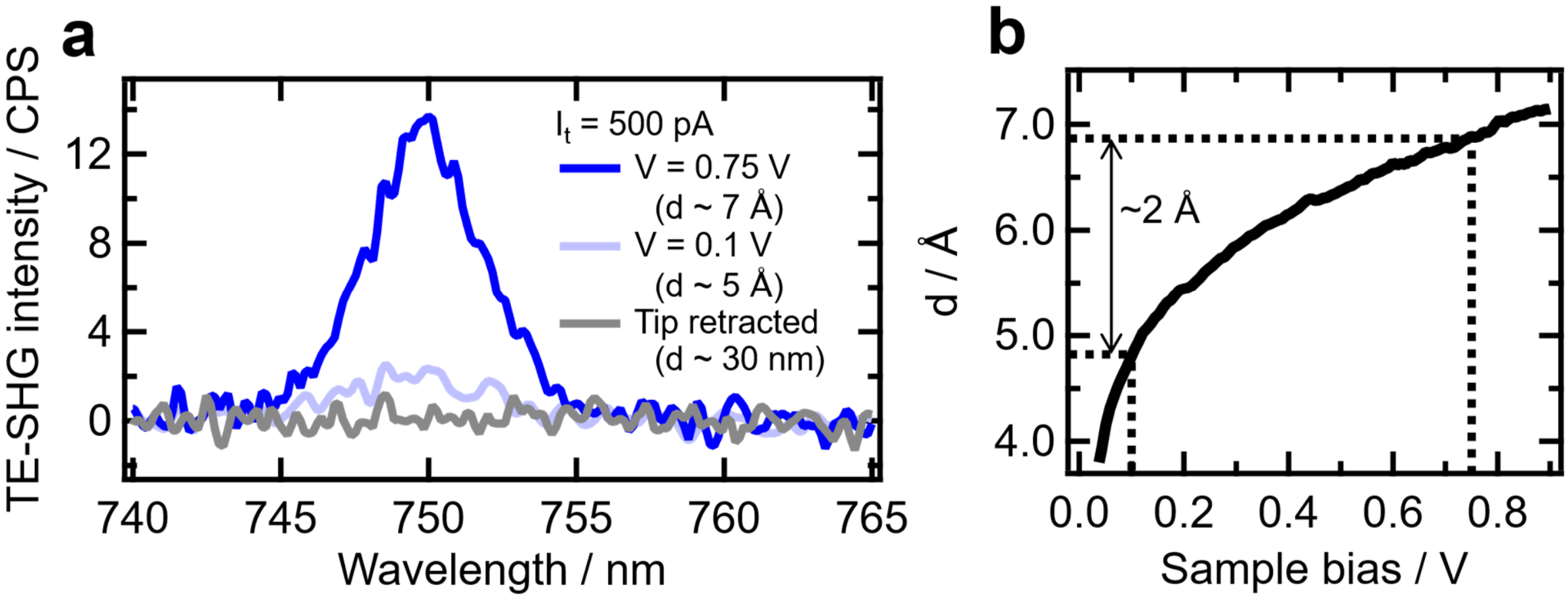

**Fig. 2 | Tip-enhanced SHG and its relationship with the applied bias and tip–substrate distance. a.** The spectra of TE-SHG excited by 1500 nm laser pulses and obtained at sample biases of 0.1 V (light blue, $d \sim 5$ Å) and 0.75 V (dark blue, $d \sim 7$ Å) with a constant tunnelling current of 500 pA. The tip–substrate distance $d$ is defined as the metal-to-metal separation between the tip apex and the substrate surface. Gray curve indicates the signal obtained when the substrate was retracted enough from the tip ($d \sim 30$ nm) to deactivate the plasmonic enhancement effects. **b.** Dependence of the tip-substrate distance $d$ on the sample bias. The measurement was performed at a constant tunneling current of 500 pA. The voltage increase from 0.1 V to 0.75 V under this constant current condition elongated the tip–substrate distance from ~5 Å to ~7 Å. For further details of the estimation of the absolute tip–substrate gap distance, see Supplementary Section 3.

Then, the tip-substrate distance ($d$), defined as the metal-to-metal separation between the tip apex and the substrate surface, was reduced from ~30 nm to ~5 Å under the sample bias of 0.1 V. In this condition, not only the intensity of the forward-scattered SHG signal increased (light blue curve in Supplementary Fig. S6), but also the backward-scattered SHG signal newly appeared (light blue curve in Fig. 2a). In this case, since the movement of the piezoelectric stage is quite small (~30 nm) compared to the optical wavelength of light, the signal collection efficiency should remain unchanged even after the formation of the angstrom-scale junction. Moreover, while the tip apex is expected to slightly indent into the SAM layer at a tip–substrate distance of ~5 Å, the potential influence of such contact, including chemical enhancement effects[52], should be negligibly small because the terminal methyl group of MBT molecules is chemically inert (further details in Supplementary Section 3). Therefore, the signal increase in the forward-scattering geometry (light blue curve in Supplementary Fig. S7) and the appearance of the backward-scattered signal (light blue curve in Fig. 2a) are obviously due to the near-field enhancement activated within the angstrom-scale tip-substrate gap, and thus the observed signals correspond to TE-SHG emission. Note that the technical basis for detecting such tip-enhanced nonlinear optical signals separately from far-field signals has already been established and reported in our recent publication[48,49].

While we also detected TE-SHG from the bare Au(111) surface without molecular adsorption (Supplementary Fig. S13), the observed intensity was much lower than that observed for the Au substrate coated with MBT SAM. We consider that the TE-SHG signal observed in the absence of the SAM is attributed to the nonlinear optical response of the surface electrons of the Au tip and substrate. Since the electric field enhancement at the central region of the gap is expected to be much stronger than that near the top surfaces of the Au substrate and Au tip[53], the SAM film placed within the gap region can produce significantly stronger TE-SHG emission than the surface electrons of Au tip and substrate. Therefore, the main source of TE-SHG signals at 0.1 V shown in Fig. 2a and Supplementary Fig. S7 are the vibrationally non-resonant $\chi^{(2)}$ response of the MBT SAM, which is amplified through the field enhancement effect caused by the tip–substrate gap (further details in Supplementary Section 10). In this case, the observed spectral shape of TE-SHG reflects that of the near-IR excitation pulses (Supplementary Fig. S2).

As also shown in Fig. 2a and Supplementary Fig. S7, we found that the TE-SHG intensity markedly increased when the sample bias ($V$) was increased from 0.1 V to 0.75 V, with the tunneling current setpoint maintained constant (500 pA). This voltage increase at a constant tunneling current (500 pA) extended the tip–sample distance from ~5 Å to ~7 Å (Fig. 2b). Based on conventional classical electrodynamic simulations, the near-field enhancement strength is expected to decrease by several tens of percents during this ~2-Å tip–substrate distance elongation[54–59]. However, at the gap distances of < 1 nm, the influences of quantum mechanical phenomena, such as electron spill-out from the metal surface and the overlap of electronic wavefunctions across the gap, begin to quench plasmon excitations and suppress the increase in field enhancement[53,56,57,60–63]. Particularly, in the gap distance range of ~4–7 Å, these quantum suppression effects and the classically expected enhancement nearly cancel each other, making the overall field enhancement effectively independent of the tip–substrate distance[53,56,57,60–63]. Therefore, the near-field enhancement strength should remain essentially unaffected during the distance elongation from 5 Å to 7 Å. Despite such constant field enhancement, the TE-SHG intensity significantly increased with the applied voltage (Fig. 2a). Importantly, this distance expansion occurred when the bias voltage was raised from 0.1 V to 0.75 V at a constant tunneling current (500 pA). With this voltage increase, the electrostatic field below the tip apex increased from $2 \times 10^8$ V/m to $1 \times 10^9$ V/m. Therefore, both the TE-SHG intensity and the electrostatic field increased concurrently, indicating the presence of substantial electriostatic field-induced effects that boost the overall TE-SHG intensity.

To gain more quantitative insights into the field-effect modulation of the TE-SHG intensity, we fixed the tip–substrate distance at $d \sim 7$ Å and monitored the voltage dependent variation in the TE-SHG intensity (Fig. 3). At this tip–substrate distance, the classically predicted near-field enhancement and the quantum plasmonic quenching effects arrive at a nearly optimal balance, resulting in almost maximum near-field enhancement values[53,56,57,60–63]. Moreover, as long as the tip–substrate distance is kept constant, variations in the STM bias voltage on the order of 1 V essentially unaffect the near-field enhancement factor within the gap[64]. Therefore, fixing the tip–substrate distance at ~7 Å can be regarded as the optimal condition under which the field enhancement can be maintained at its nearly maximum value regardless of the applied bias voltage. This enables us to exclusively examine the influence of electrostatic fields on the TE-SHG process, while eliminating the contributions of variations in the near-field enhancement. Note that the constant tip–substrate distance of 7 Å was kept by adjusting the tunneling current setpoint synchronously with the voltage sweep (further details in Supplementary Section 4). During the voltage sweep, the excitation intensity of the incident laser was maintained low and constant at 0.5 mW. Despite such fixed tip–substrate distance and excitation power, the TE-SHG intensity significantly grew up when increasing the sample bias voltage from 0.1 V to 1 V. Importantly, such a significant TE-SHG signal modulation was also observed when we swept the bias voltage inversely from $-0.1$ V to $-1$ V, exhibiting a quadratic dependence on the applied voltage (Fig. 3). More remarkably, the change in the TE-SHG signal output ($\Delta I(V)$) relative to the signal at $V = 0$ V ($I(V = 0)$) reached ~2000% at $V = \pm 1$ V (Fig. 3). The stability and reproducibility of these results were confirmed through repeated measurements as described in detail in Supplementary Sections 6 and 12. Moreover, similar voltage-controlled modulation of TE-SHG was also observed under the constant tunneling current conditions (Supplementary Fig. S14), indicating that the contribution of tunneling current to the observed variations in the TE-SHG intensity (Fig. 3) is negligible. Therefore, we can reasonably conclude that the observed electrical modulation behavior is predominantly governed by voltage-induced electrophotonic effects, distinct from variations in the near-field enhancement or contributions from the tunneling current.

Notably, the observed near-field nonlinear electrophotonic effects, characterized by a quadratic response and a large modulation depth of ~2000%/V, are in stark contrast with previous studies[42–46]. Over the past decade, electric-field modulations in plasmon-enhanced near-field nonlinear optics have been pioneered with sub-100-nm scale gap structures, within which organic molecules or inorganic insulators were embedded as nonlinear optical media. However, due to the inherent challenges in fabricating angstrom-scale gaps, the sub-100-nm scale structures used in these previous studies exhibited modulation depths of the order of 10 %/V, with a linear response to the applied voltage[42–46]. In contrast, our angstrom-scale plasmonic gap occupied by a molecular SAM exhibits a quadratic voltage response with an unprecedented electrical modulation depth of 2000%/V. Moreover, we confirmed that the surface electrons of gold also generate TE-SHG signals even without the SAM and exhibits a similar quadratic voltage dependence with an electrical modulation depth of ~1000%/V (Supplementary Fig. S13). We consider that the giant TE-SHG modulation observed for the bare Au(111) surface (Supplementary Fig. S13) is likely attributed to the electric field-induced change in the free electron density at the topmost Au surface[65]. These results underscore that the angstrom-scale plasmonic gap serves as a medium-independent platform for achieving substantial nonlinear optical modulation depths of at least ~1000%/V with a modest bias of ≤ 1 V.

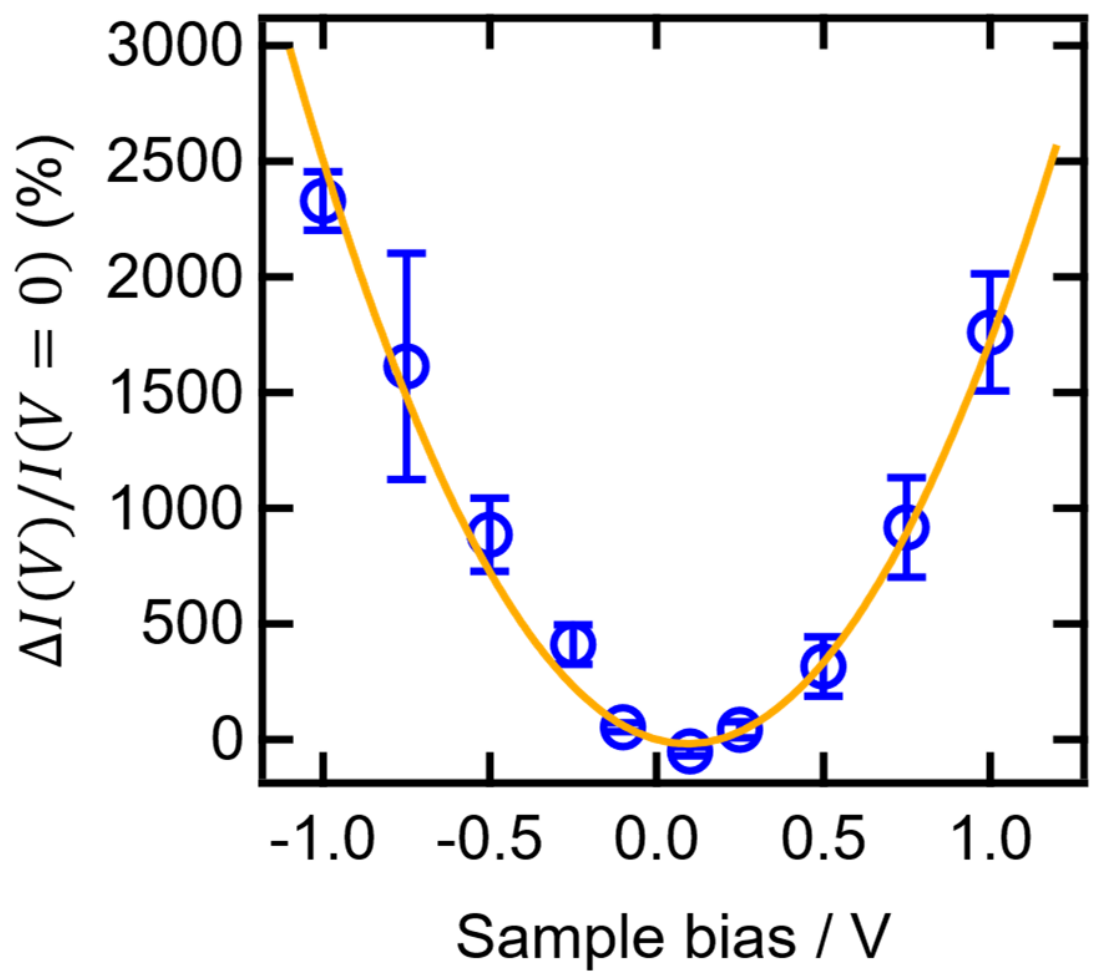

**Fig. 3 | Giant electric modulation of TE-SHG.** The voltage-dependent change in the TE-SHG intensity ($\Delta I(V)$) normalized by the signal intensity at $V = 0$ V ($I(V = 0)$) is depicted. $I(V = 0)$ was obtained by averaging the intensities at $V = -0.1$ V and $V = 0.1$ V. The modulation depth of ~2000% is achieved at a bias voltage of ±1 V. The orange curve is the result of curve fitting with a quadratic function. The measurements were performed in the backward-scattering geometry under constant excitation intensity (0.5 mW) and constant tip–substrate distance (~7 Å). The error bars in the plot represent the standard error of each data point, which was determined by repeating the same voltage-dependent experiment three times.

Given that no structural changes were observed in the MBT SAM samples (Supplementary Fig. S12), the contribution of field-induced phase transitions or polarization switching phenomena[66–69] can be considered insignificant. Consequently, the observed gigantic modulation of TE-SHG signals is attributed primarily to electro-optic interactions triggered by the electrostatic field within the plasmonic gap ($E_{\mathrm{DC}}$). Such electro-optic interaction in the second-order SHG output is likely to be induced by the third-order nonlinear effect of molecules, representing the mixed interaction between the incident light $E_{\mathrm{gap}}(\omega)$ and the electrostatic field $E_{\mathrm{DC}}$, $P^{(3)}(2\omega) = \varepsilon_0 \chi^{(3)}(2\omega; 0, \omega, \omega) E_{\mathrm{DC}} E_{\mathrm{gap}}(\omega) E_{\mathrm{gap}}(\omega)$. In this case, the total TE-SHG signal intensity ($I_{\mathrm{TESHG}}$) includes contributions not only from the second-order nonlinear polarization ($P^{(2)}(2\omega) = \varepsilon_0\chi^{(2)}(2\omega;\omega,\omega)E_{\mathrm{gap}}(\omega)E_{\mathrm{gap}}(\omega)$) but also from the electrically tunable third-order nonlinear polarization ($P^{(3)}(2\omega)$), as follows:

$$I_{\mathrm{TESHG}} \propto \left|\chi^{(2)} + \chi^{(3)}E_{\mathrm{DC}}\right|^2 {I_{\mathrm{gap}}}^2, \tag{1}$$

where $I_{\mathrm{gap}}$ denotes the intensity of the tip-enhanced electric field of the incident excitation pulses $\left(I_{\mathrm{gap}} \propto \left|E_{\mathrm{gap}}(\omega)\right|^2\right)$; and $\chi^{(2)}$ and $\chi^{(3)}$ are second- and third-order nonlinear optical susceptibility, respectively, of the medium sensing the enhanced-electric field in the angstrom-scale gap. Note that our recent work demonstrated that, when the tip apex size is on the order of ~50 nm as in the present study (Fig. 1b), the generated nonlinear polarizations are predominantly dipolar in nature, with negligible contributions from quadrupolar or higher-order multipolar components[70]. Moreover, $E_{\mathrm{gap}}$ and $E_{\mathrm{DC}}$ in eq. (1) represent the spatially averaged field values over the ~20-nm field enhancement region beneath the tip apex (see Supplementary Section 14 for details). Since these intragap fields are dominated by surface-normal (*Z*-directed) components (Supplementary Figs. S19 and S20), among the multiple tensor elements of $\chi^{(2)}$ and $\chi^{(3)}$, only $\chi^{(2)}_{ZZZ}$ and $\chi^{(3)}_{ZZZZ}$ components dominantly contribute to the generation of TE-SHG signals (further details in Supplementary Section 14). The incorporation of the field-induced $\chi^{(3)}E_{\mathrm{DC}}$ term in the equation (1) is referred to as the electric-field-induced second-harmonic (EFISH) effect. Notably, during the bias-dependent TE-SHG measurements (Fig. 3), the excitation intensity was fixed at 0.5 mW, and the field enhancement strength within the gap was kept constant by maintaining a fixed tip–substrate distance (7 Å). Under these controlled conditions, $I_{\mathrm{gap}}$ in eq. (1) can be regarded as constant throughout the bias-dependent TE-SHG experiments. Therefore, the observed substantial bias dependence (Fig. 3) represents the first experimental achievement in observing an unprecedentedly giant near-field EFISH effect arising from the $\chi^{(3)}E_{\mathrm{DC}}$ term in an angstrom-scale plasmonic gap structure, with sufficient stability and reproducibility (Supplementary Sections 6 and 12).

By expanding the right-hand side of the equation (1), we obtain a voltage-independent term $\left(I_{\mathrm{TESHG}}(V=0) = \left|\chi^{(2)}\right|^2 {I_{\mathrm{gap}}}^2\right)$ and voltage-dependent terms $\left(\Delta I_{\mathrm{TESHG}}(V) = \left[\left|\chi^{(3)}\right|^2 E_{\mathrm{DC}}^2 + 2\mathrm{Re}\left(\chi^{(3)}\chi^{(2)*}\right)E_{\mathrm{DC}}\right] {I_{\mathrm{gap}}}^2\right)$. When the contribution from $\chi^{(3)}E_{\mathrm{DC}}$ is significantly smaller than that from $\chi^{(2)}$, corresponding to the modest electrical modulation depth on the order of at most 10 %/V, the voltage-dependent component is dominated by the linear $E_{\mathrm{DC}}$ term[42–46,71,72], resulting in limited modulation. In contrast, when the $\chi^{(3)}E_{\mathrm{DC}}$ term far exceeds the $\chi^{(2)}$ term, corresponding to much larger modulation depth exceeding 1000 %/V, the voltage-dependent component becomes dominated by the quadratic $E_{\mathrm{DC}}$ term, leading to significantly enhanced modulation. Previous studies employing sub-100-nm plasmonic gap structures generated relatively modest electric fields (~$10^7$ V/m), resulting in linear voltage dependence with modest modulation depths (~10%/V)[42–46,71,72]. In this case, a high bias on the order of 100 V is required to achieve a modulation depth of approximately 1000%. In contrast, our approach leveraged an angstrom-scale plasmonic gap ($d < 10$ Å), facilitating a much more intense electrostatic field (~$10^9$ V/m) even under a moderate voltage application (~1 V), within a range that does not damage metal electrodes or molecules[73–75]. This resulted in a dominant contribution from the $\chi^{(3)}E_{\mathrm{DC}}$ term, achieving a significantly larger modulation depth on the order of 1000%/V with a quadratic voltage dependence. These findings highlight a pronounced near-field nonlinear electrophotonic effect unique to angstrom-scale gap structures. By fitting the bias-dependent intensity change in Fig. 3, and assuming $d = 7$ Å, the relative values of $\chi^{(2)}$ and $\chi^{(3)}$ were derived as $|\chi^{(3)}|^2/|\chi^{(2)}|^2 = (10.3 \pm 0.4)\ \mathrm{nm}^2/\mathrm{V}^2$ and $2\mathrm{Re}(\chi^{(3)}\chi^{(2)*})/|\chi^{(2)}|^2 = (-2.7 \pm 0.5)\ \mathrm{nm/V}$. From these relative values, we can estimate the threshold field ($E_{\mathrm{DC,0}}$), at which $\chi^{(2)}$ and $\chi^{(3)}E_{\mathrm{DC}}$ effects equally contributes to the nonlinear optical generation $\left(|\chi^{(3)}E_{\mathrm{DC,0}}| = |\chi^{(2)}|\right)$, to be $3.1 \times 10^8$ V / m. Assuming the gap distance of $d = 7$ Å, the corresponding applied voltage is estimated to be ± 0.22 V. Thus, when applied voltage exceeds this value, the contributions from $\chi^{(3)}E_{\mathrm{DC}}$ becomes dominant over that from $\chi^{(2)}$.

It should be noted that although sub-angstrom (~0.5–1 Å) fluctuations in $d$, which should induce variations in $E_{\mathrm{DC}}$, are inevitably present under our ambient experimental conditions (Supplementary Fig. S11d), their influence was not incorporated into the fitting analysis. This is because the typical timescale of the fluctuations is on the order of microsecond and thus much shorter than minute-scale signal integration time in our measurements. Consequently, the influence of the fluctuations in $E_{\mathrm{DC}}$ would be effectively averaged out and would not manifest in the measured modulation curves. For this reason, the fluctuations in $E_{\mathrm{DC}}$ were not explicitly taken into account in our fitting analysis.

**Extension to tip-enhanced sum-frequency generation.** Nonlinear optical processes generally involve multiphoton interaction and are often accompanied by drastic frequency conversion between the incoming and outgoing light. We recently demonstrated that such multicolored input and output light with separated frequencies can be simultaneously enhanced in a tip–substrate gap structure, thereby enabling the efficient plasmonic enhancement of nonlinear optical processes involving significant wavelength conversion[48] (further details in Supplementary Section 13). The TE-SHG process discussed thus far exemplifies this phenomenon, because it arises from the simultaneous enhancement of both near-IR excitation and visible emission[48]. While the giant nonlinear electrophotonic modulation has been demonstrated for this near-IR-to-visible TE-SHG process (Fig. 3), extending this modulation to other nonlinear optical processes involving even more pronounced wavelength conversion would further substantiate the broadband applicability of the observed nonlinear electrophotonic effects. In this context, we next focus on voltage dependence of a tip-enhanced SFG (TE-SFG) process,[49] another representative second-order nonlinear optical process characterized by a large frequency upconversion from the mid-IR input into near-IR or even visible output (Fig. 4a).

For the demonstration of voltage-dependent TE-SFG measurements in the angstrom-scale plasmonic junctions (Fig. 4b), we used spatially and temporarily overlapped mid-IR ($\omega_1$: 3280 nm, 300 fs, FWHM: 60 nm, 20 mW) and near-IR ($\omega_2$: 1033 nm, 1 ps, FWHM: 1 nm, 0.5 mW) pulses as the excitation light with a repetition frequency of 50 MHz, both detuned from molecular resonance. Moreover, while the STM chamber was maintained at ultra-high vacuum condition in the TE-SHG measurements, we exposed the chamber to the air atmosphere prior to the TE-SFG experiments in order to verify the operability of nonlinear electrophotonic modulation in open air under ambient conditions. Then, as shown in Fig. 4c, the vibrationally non-resonant TE-SFG signals were clearly observed in the visible region around 785 nm. The observed spectral shape and band width predominantly reflect those of the incident mid-IR excitation pulse (Supplementary Fig. S3). Similar to the results of TE-SHG experiments in Fig. 2a, the TE-SFG intensity also increased with the sample bias voltage at constant current of 250 pA (Fig. 4c). While the tip–substrate distance varied from 6.7 Å to 8.2 Å in this experiment, the distance dependence of the electric field enhancement strength should be negligibly small due to the quantum plasmonic quenching effects[53,56,57,60–63]. Therefore, by analogy with the case of the bias-dependent TE-SHG (Fig. 2a), the observed TE-SFG intensity modulation shown in Fig. 4c can also be attributed to bias-induced effects.

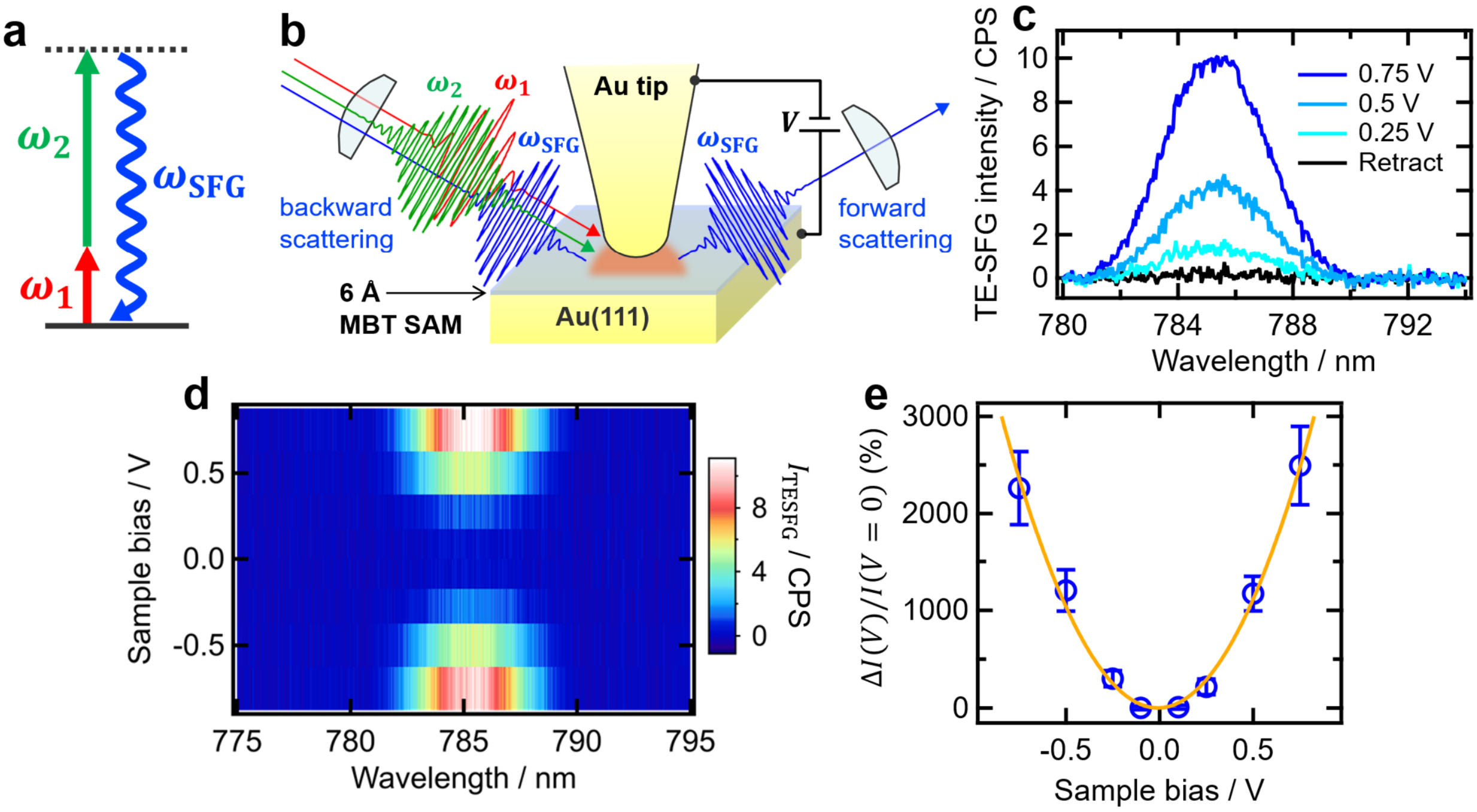

**Fig. 4 | Giant electric modulation of TE-SFG. a.** Energy level diagram of SFG process. Two-photon

excitation with mid-IR ($\omega_1$) and near-IR ($\omega_2$) light induces radiation at the sum frequency of those light ($\omega_{\mathrm{SFG}} = \omega_1 + \omega_2$) **b.** Schematic depiction of the TE-SFG experiment using a Au tip and a Au substrate with an MBT SAM mounted on an STM unit. The experiments were conducted under ambient pressure in an air atmosphere at room temperature. **c.** TE-SFG spectra obtained at sample biases of 0.25 V (cyan, $d$~6.7 Å), 0.5 V (sky blue, $d$~7.6 Å), and 0.75 V (dark blue, $d$~8.2 Å) with a constant tunnelling current setpoint of 250 pA. The black curve indicates the signal obtained when the tip and the substrate were retracted enough (30 nm) to deactivate the plasmonic enhancement effects. The excitation intensity of mid-IR ($\omega_1$) and near-IR ($\omega_2$) was 20 mW and 0.5 mW, respectively. **d.** Two-dimensional plot of TE-SFG spectra obtained with the various applied sample bias under the almost constant tip–substrate distance ($d$~7 Å). **e.** The voltage-dependent change in the TE-SFG output ($\Delta I(V)$) normalized by the TE-SFG signal intensity at $V = 0$ V ($I(V = 0)$). The orange curve is the result of curve fitting with a quadratic function. The error bars in the plot represent the standard error of each data point, which was determined by repeating the same voltage-dependent experiments ten times.

Then, we investigated the impact of bias voltage while maintaining an almost constant tip–substrate distance ($d$~7 Å). As shown in Fig. 4d, the intensity of the TE-SFG signals drastically increased with increasing the applied voltage in both positive (0.1 V → 1 V) and negative (−0.1 V → −1 V) directions. Consistent with the results of electrically controlled TE-SHG experiments (Fig. 3), $\Delta I(V)$ exhibited a quadratic dependence on the applied voltage, with the intensity enhancement at $V \approx \pm 1$ V exceeding 2000% of the TE-SFG intensity at $V = 0$ V ($I(V = 0)$) (Fig. 4e). Similar bias dependences were also obtained using other Au tips (Supplementary Fig. S15), ensuring the reproducibility of our experiments. These findings demonstrate that the giant electric-field induced modulation mechanism of plasmonic nonlinear optical effects is also highly effective for TE-SFG ( $I_{\mathrm{TESFG}} \propto \left|\chi^{(3)} E_{\mathrm{DC}} + \chi^{(2)}\right|^2 I(\omega_1) I(\omega_2) \cong \left|\chi^{(3)} E_{\mathrm{DC}}\right|^2 I(\omega_1) I(\omega_2)$ ) and operates over a broadband wavelength range, encompassing the mid-IR, near-IR, and visible regions. This broad-wavelength applicability is in stark contrast to the previously reported electronic resonance-assisted EFISH study[76], where the range of excitation wavelength that can be used to detect the large EFISH effect was significantly restricted. Note that the slight difference in the minima of the TE-SHG (Fig. 3) and TE-SFG (Fig. 4e) curves may arise from minor variations in the atomic-scale shape of the Au tips and the resultant difference in the electrostatic field distribution within the gap.

Remarkably, the achievement of more than 2000%/V modulation of TE-SFG signals in an angstrom-scale plasmonic gap (Fig. 4e) was realized under room temperature and ambient pressure conditions, rather than low temperature and ultra-high vacuum environments. Even under these ambient conditions, the stability of the signals under light irradiation (Supplementary Section 6) and the reproducibility of the results with different STM tips (Supplementary Fig. S15) were thoroughly verified. In addition, the possibility of the dielectric breakdown caused by an intense electrostatic field within the gap (~$10^9$ V/m) can also be reasonably ruled out in our ambient experimental conditions, because of the extremely rare electron–gas collision within our angstrom-scale gap (further details in Supplementary Section 9). This allows us to safely disregard any intense-field-induced modification of the tip or substrate even under ambient experimental conditions. Furthermore, the near-field tip-enhancement scheme constructed under room temperature achieves markedly higher SHG and SFG conversion efficiencies than conventional far-field excitation, despite the much smaller number of molecules involved in the signal generation process in the near-field scheme (further details in Supplementary Section 7). These findings highlight the promising potential of the observed nonlinear electoplasmonic phenomena and pave the way for the development of photonic devices that combine substantial modulation depth, operation under ambient conditions, and higher conversion efficiency.

Finally, it should be noted that our near-field-based giant electrical modulation depth on the order of 2000%/V compares favorably with, and even exceeds, the recently reported record-high values (>1000%/V) for practical applications in the field of far-field-based electrophotonics[66–69,76]. In far-field schemes, the electric field across macroscopic electrodes is typically much weaker than that in near-field schemes. Such weaker electrostatic fields alone are generally insufficient to induce giant electro-optic interactions. Consequently, the record-high values in far-field approaches have been achieved primarily through some medium-design strategies. One approach was to minimize the contribution of $\chi^{(2)}$ with respect to $\chi^{(3)}$ using centrosymmetric amorphous material films with ~100 nm thickness[76]. Another important approach was to utilize field-induced phase transition or polarization switching phenomena in specific electroactive nonlinear optical media, such as transition metal dichalcogenides[66], superlattice structures[67,68], and ferroelectric sheets[69]. In contrast to these approaches that rely on specialized material designs, our near-field-based giant electrical modulation was achieved using a common SAM of electro-nonactive organic molecules that do not exhibit field-induced polarization switching and drastic phase transition. Furthermore, a similar extent of electric modulation (~1000%/V) was also observed even in the absence of the SAM (Supplementary Fig. S13). These results demonstrate that, in our near-field-based nonlinear optical modulation scheme, the choice of material is largely noncritical for achieving modulation depth of at least 1000%/V. Therefore, the angstrom-scale plasmonic gaps have great potential to serve as an optimal platform for giant nonlinear optical modulation. A promising avenue for further exploration is the use of more electroactive nonlinear optical media, similar to those used in far-field based approaches[66–69], to further improve modulation performance. Such advancements could achieve practical level of modulation depth (~1000 %) even under more modest bias voltages, paving the way for an atomic-scale electrophotonic information processing technology.

## Conclusion

By combining the STM and wavelength tunable pulse laser system, we have demonstrated unprecedentedly giant electric enhancement of plasmonic near-field SHG and SFG responses within an angstrom-scale plasmonic gap under room temperature, achieving a remarkable modulation depth of ~2000%/V not only at ultra-high vacuum conditions but also in open air under ambient conditions. The observed quadratic bias dependence and significant signal modulation are in stark contrast to the quasi-linear bias dependence and modest modulation depths (~10%/V) previously reported in nanoscale plasmonic gap systems[42–46]. Therefore, our findings lay the foundation for a paradigm shift from sub-100-nanometer to angstrom-scale electrophotonics enabling ultrahigh-performance nonlinear optical modulation. We also identified several key advantages of angstrom-scale plasmonic structures: moderate driving voltage (<1 V), broad material compatibility without requiring specific electroactive media, operability under ambient conditions, and a broad operation wavelength range spanning the visible, near-IR, and mid-IR regions. The realization of electrophotonic systems that integrate all these properties into an ultra-compact platform far below the diffraction limit is a pivotal step toward the new era of atomic-scale electrophotonics. We believe that our results and concepts not only demonstrate the potential of near-field-based nonlinear electrophotonic modulation but also establish an experimental basis for integrating nanophotonics with atomic-scale electronics[77–79], driving further optimization of device architectures and advancements in angstrom-level photonic information processing techniques.

## Methods

**Preparation of STM tip and sample.** Gold tips for STM were prepared by electrochemical etching of gold wires with a diameter of 0.25 mm (Nilaco) in a 2.79 M KCl (FUJIFILM Wako Pure Chemical Corporation) aqueous solution[47]. After the etching process completed, the prepared tips were washed by concentrated $H_2SO_4$ and ultrapure water and introduced into our STM chamber (USM1400, UNISOKU). Prior to the nonlinear optical measurements, we processed the Au tips with $Ar^+$ sputtering ($2\times10^{-3}$ Pa, 3 kV) for 3 hours.

For the substrate to form the angstrom-scale gap structure, a 200-nm-thick gold thin film vapor-deposited on a mica substrate (UNISOKU) was employed. The substrate was first annealed in a butane flame to achieve an atomically flat Au(111) surface (Supplementary Fig. S1a) and cooled down to room temperature. Then, the Au substrate was immersed in 1 mM ethanolic solution of 4-Methylbenzenethiol (MBT) at room temperature for 48 hours to form a ~6-Å-thick self-assembled monolayer (SAM) of MBT on its surface (Supplementary Fig. S1b)[51]. Thereafter, the Au substrate was picked up from the solution, rinsed with pure ethanol, and dried with $N_2$ gas just before being placed into our STM chamber.

**Optical system for tip-enhanced nonlinear optical measurements.** All the nonlinear optical measurements were performed at room temperature. Our optical system is based on an amplified Yb-fiber laser (1033 nm, 280 fs, 40 W, 50 MHz; Monaco-1035-40-40, Coherent). The fundamental output from the laser was divided into two portions by a beam splitter. The first portion (120 nJ) was used to drive a commercially available optical parametric oscillator (OPO, Levante IR, APE), generating two kinds of wavelength-tunable IR pulses: signal (1.3 – 2 μm) and idler (2.1 – 5 μm). The second portion was passed through an air-spaced Fabry-Pérot etalon to narrow down the spectral width. Three kinds of pulses (signal, idler, and narrow-band fundamental wave) were used as excitation sources of TE-SHG and TE-SFG experiments, as described below. See Supplementary Figs. S2 and S3 for the spectra of those pulses.

For TE-SHG experiments, the STM chamber was maintained at ultra-high vacuum condition with a base pressure of $<1\times10^{-7}$ Pa. The OPO's signal output centered at 1500 nm was used as an excitation source. The excitation light intensity was attenuated to 0.5 mW (pulse energy: 10 pJ, peak power density: $5 \times 10^6$ W/cm$^2$) by a variable neutral density filter, and then focused onto the apex of the STM tip at an incident angle of 55° using a custom-made $CaF_2$ aspherical lens (NA = 0.34, focal length of 19.3 mm at 1033 nm) installed on the STM head. The incident light was linearly polarized along the axis of the tip ($p$-polarized). The backward-scattered TE-SHG signal at 750 nm was collimated with the same lens and separated from the incident light path using a custom-made dichroic mirror. The forward-scattered TE-SHG signal was collimated with another lens mounted in the reflection direction and then the incident light was cut off by using a short pass filter. Those two kinds of signals were separately input into a bundle fiber branched into two halves without polarization selection and directed into a spectrometer (Kymera 328i, Andor). Within the spectrometer, the signals from both directions were focused on the distinct vertical positions on the same electronically cooled CCD detector (iDus 416, Andor), allowing us to separately measure those two signals at the same time. The spectrometer wavelength was calibrated using a Hg/Ar lamp (SL2, StellarNet). When measuring voltage-dependent TE-SHG signals shown in Fig. 3, the measurements were repeated three times, with the order of the voltage sweeps varied for each measurement (further details in Supplementary Section 6), and the signal accumulation time for each voltage point was 10 seconds.

For TE-SFG experiments, we extended this optical setup to a two-color excitation scheme and the STM chamber was maintained at ambient pressure in an air atmosphere. To drive TE-SFG process, the near-IR fundamental output from the Yb-fiber laser (0.5 mW, pulse energy: 10 pJ, peak power density: $5 \times 10^6$ W/cm$^2$) and the mid-IR idler output (20 mW, pulse energy: 400 pJ, peak power density: $2 \times 10^6$ W/cm$^2$) were spatially and temporarily overlapped. The two laser pulses were combined collinearly at a dichroic mirror (reflective for 1033 nm and transparent for 1300–7000 nm) and then focused onto the apex of the STM tip by the same aspherical lens used in TE-SHG measurements. Both excitation pulses were linearly polarized along the axis of the tip (*p*-polarized). The detection of the signals was performed using the same method as that used for TE-SHG measurements. When measuring voltage-dependent TE-SFG signals shown in Figs. 4d and e, the applied voltages were varied in a random order, and the signal accumulation time for each voltage point was 300 seconds (further details in Supplementary Section 6).

**Data availability.** All experimental data in this study are available from the corresponding author upon reasonable request.

## References

1. Wang, X. *et al.* Quantum frequency conversion and single-photon detection with lithium niobate nanophotonic chips. *npj Quantum Inf.* **9**, 1–7 (2023).
2. Sasikala, V. & Chitra, K. All optical switching and associated technologies: a review. *J. Opt.* **47**, 307–317 (2018).
3. Débarre, D. *et al.* Imaging lipid bodies in cells and tissues using third-harmonic generation microscopy. *Nat. Methods* **3**, 47–53 (2006).
4. Xu, P. *et al.* Cation Modifies Interfacial Water Structures on Platinum during Alkaline Hydrogen Electrocatalysis. *J. Am. Chem. Soc.* **146**, 2426–2434 (2024).
5. Nihonyanagi, S., Yamaguchi, S. & Tahara, T. Ultrafast Dynamics at Water Interfaces Studied by Vibrational Sum Frequency Generation Spectroscopy. *Chem. Rev.* **117**, 10665–10693 (2017).
6. Kauranen, M. & Zayats, A. V. Nonlinear plasmonics. *Nat. Photon.* **6**, 737–748 (2012).
7. Hasan, S. B., Lederer, F. & Rockstuhl, C. Nonlinear plasmonic antennas. *Mater. Today* **17**, 478–485 (2014).
8. Butet, J., Brevet, P.-F. & Martin, O. J. F. Optical Second Harmonic Generation in Plasmonic Nanostructures: From Fundamental Principles to Advanced Applications. *ACS Nano* **9**, 10545–10562 (2015).
9. Panoiu, N. C., Sha, W. E. I., Lei, D. Y. & Li, G.-C. Nonlinear optics in plasmonic nanostructures. *J. Opt.* **20**, 083001 (2018).
10. Ron, R., Zar, T. & Salomon, A. Linear and Nonlinear Optical Properties of Well-Defined and Disordered Plasmonic Systems: A Review. *Adv. Opt. Mater.* **11**, 2201475 (2023).
11. Aouani, H., Rahmani, M., Navarro-Cía, M. & Maier, S. A. Third-harmonic-upconversion enhancement from a single semiconductor nanoparticle coupled to a plasmonic antenna. *Nat. Nanotechnol.* **9**, 290–294 (2014).

12. Vampa, G. *et al.* Plasmon-enhanced high-harmonic generation from silicon. *Nat. Phys.* **13**, 659–662 (2017).
13. Huang, L. *et al.* Surface plasmon enhanced THz emission with nanoporous gold supported CdTe. *Opt. Express* **29**, 19853–19861 (2021).
14. Mesch, M., Metzger, B., Hentschel, M. & Giessen, H. Nonlinear Plasmonic Sensing. *Nano Lett.* **16**, 3155–3159 (2016).
15. Ghirardini, L. *et al.* Plasmon-Enhanced Second Harmonic Sensing. *J. Phys. Chem. C* **122**, 11475–11481 (2018).
16. Verma, M. S. & Chandra, M. Nonlinear Plasmonic Sensing for Label-Free and Selective Detection of Mercury at Picomolar Level. *ACS Sens.* **5**, 645–649 (2020).
17. Verma, M. S. & Chandra, M. Second harmonic generation-based nonlinear plasmonic RI-sensing in solution: the pivotal role of the particle size. *Phys. Chem. Chem. Phys.* **23**, 25565–25571 (2021).
18. Wang, C.-F. & El-Khoury, P. Z. Multimodal (Non)Linear Optical Nanoimaging and Nanospectroscopy. *J. Phys. Chem. Lett.* **13**, 7350–7354 (2022).
19. Wang, C.-F. & El-Khoury, P. Z. Multimodal Tip-Enhanced Nonlinear Optical Nanoimaging of Plasmonic Silver Nanocubes. *J. Phys. Chem. Lett.* **12**, 10761–10765 (2021).
20. Wang, C.-F. & El-Khoury, P. Z. Imaging Plasmons with Sub-2 nm Spatial Resolution via Tip-Enhanced Four-Wave Mixing. *J. Phys. Chem. Lett.* **12**, 3535–3539 (2021).
21. Park, K.-D. & Raschke, M. B. Polarization Control with Plasmonic Antenna Tips: A Universal Approach to Optical Nanocrystallography and Vector-Field Imaging. *Nano Lett.* **18**, 2912–2917 (2018).
22. Kravtsov, V., Ulbricht, R., Atkin, J. M. & Raschke, M. B. Plasmonic nanofocused four-wave mixing for femtosecond near-field imaging. *Nat. Nanotechnol.* **11**, 459–464 (2016).
23. Ichimura, T., Hayazawa, N., Hashimoto, M., Inouye, Y. & Kawata, S. Tip-Enhanced Coherent Anti-Stokes Raman Scattering for Vibrational Nanoimaging. *Phys. Rev. Lett.* **92**, 220801 (2004).
24. Furusawa, K., Hayazawa, N., Catalan, F. C., Okamoto, T. & Kawata, S. Tip-enhanced broadband CARS spectroscopy and imaging using a photonic crystal fiber based broadband light source. *J. Raman Spectrosc.* **43**, 656–661 (2012).
25. Luo, Y. *et al.* Imaging and controlling coherent phonon wave packets in single graphene nanoribbons. *Nat. Commun.* **14**, 3484 (2023).
26. Campos, A. *et al.* Plasmonic quantum size effects in silver nanoparticles are dominated by interfaces and local environments. *Nat. Phys.* **15**, 275–280 (2019).
27. Ringe, E. *et al.* Plasmon Length: A Universal Parameter to Describe Size Effects in Gold Nanoparticles. *J. Phys. Chem. Lett.* **3**, 1479–1483 (2012).

28. Henry, A.-I. *et al.* Correlated Structure and Optical Property Studies of Plasmonic Nanoparticles. *J. Phys. Chem. C* **115**, 9291–9305 (2011).
29. Rossner, C., König, T. A. F. & Fery, A. Plasmonic Properties of Colloidal Assemblies. *Adv. Opt. Mater.* **9**, 2001869 (2021).
30. Kilic, U. *et al.* Broadband Enhanced Chirality with Tunable Response in Hybrid Plasmonic Helical Metamaterials. *Adv. Funct. Mater.* **31**, 2010329 (2021).
31. Kılıç, U. *et al.* Tunable plasmonic resonances in Si-Au slanted columnar heterostructure thin films. *Sci. Rep.* **9**, 71 (2019).
32. Dionne, J. A., Diest, K., Sweatlock, L. A. & Atwater, H. A. PlasMOStor: A Metal−Oxide−Si Field Effect Plasmonic Modulator. *Nano Lett.* **9**, 897–902 (2009).
33. Kim, J. *et al.* Electrical Control of Optical Plasmon Resonance with Graphene. *Nano Lett.* **12**, 5598–5602 (2012).
34. Li, W., Zhou, Q., Zhang, P. & Chen, X.-W. Direct Electro Plasmonic and Optic Modulation via a Nanoscopic Electron Reservoir. *Phys. Rev. Lett.* **128**, 217401 (2022).
35. Alonso Calafell, I. *et al.* Giant enhancement of third-harmonic generation in graphene–metal heterostructures. *Nat. Nanotechnol.* **16**, 318–324 (2021).
36. Marinica, D. C. *et al.* Active quantum plasmonics. *Sci. Adv.* **1**, e1501095 (2015).
37. Ma, R.-M., Yin, X., Oulton, R. F., Sorger, V. J. & Zhang, X. Multiplexed and Electrically Modulated Plasmon Laser Circuit. *Nano Lett.* **12**, 5396–5402 (2012).
38. Deeb, C. & Pelouard, J.-L. Plasmon lasers: coherent nanoscopic light sources. *Phys. Chem. Chem. Phys.* **19**, 29731–29741 (2017).
39. Liang, Y., Li, C., Huang, Y.-Z. & Zhang, Q. Plasmonic Nanolasers in On-Chip Light Sources: Prospects and Challenges. *ACS Nano* **14**, 14375–14390 (2020).
40. Ayata, M. *et al.* High-speed plasmonic modulator in a single metal layer. *Science* **358**, 630–632 (2017).
41. Haffner, C. *et al.* Low-loss plasmon-assisted electro-optic modulator. *Nature* **556**, 483–486 (2018).
42. Cai, W., Vasudev, A. P. & Brongersma, M. L. Electrically Controlled Nonlinear Generation of Light with Plasmonics. *Science* **333**, 1720–1723 (2011).
43. Ding, W., Zhou, L. & Chou, S. Y. Enhancement and Electric Charge-Assisted Tuning of Nonlinear Light Generation in Bipolar Plasmonics. *Nano Lett.* **14**, 2822–2830 (2014).
44. Kang, L. *et al.* Electrifying photonic metamaterials for tunable nonlinear optics. *Nat. Commun.* **5**, 4680 (2014).
45. Lan, S., Rodrigues, S., Cui, Y., Kang, L. & Cai, W. Electrically Tunable Harmonic Generation of Light from Plasmonic Structures in Electrolytes. *Nano Lett.* **16**, 5074–5079 (2016).

46. Agreda, A. *et al.* Electrostatic Control over Optically Pumped Hot Electrons in Optical Gap Antennas. *ACS Photonics* **7**, 2153–2162 (2020).
47. Yang, B., Kazuma, E., Yokota, Y. & Kim, Y. Fabrication of Sharp Gold Tips by Three-Electrode Electrochemical Etching with High Controllability and Reproducibility. *J. Phys. Chem. C* **122**, 16950–16955 (2018).
48. Takahashi, S., Sakurai, A., Mochizuki, T. & Sugimoto, T. Broadband Tip-Enhanced Nonlinear Optical Response in a Plasmonic Nanocavity. *J. Phys. Chem. Lett.* **14**, 6919–6926 (2023).
49. Sakurai, A., Takahashi, S., Mochizuki, T. & Sugimoto, T. Tip-Enhanced Sum Frequency Generation for Molecular Vibrational Nanospectroscopy. *Nano Lett.* **25**, 6390–6398 (2025).
50. Mahapatra, S., Li, L., Schultz, J. F. & Jiang, N. Methods to fabricate and recycle plasmonic probes for ultrahigh vacuum scanning tunneling microscopy-based tip-enhanced Raman spectroscopy. *J. Raman Spectrosc.* **52**, 573–580 (2021).
51. Seo, K. & Borguet, E. Potential-Induced Structural Change in a Self-Assembled Monolayer of 4-Methylbenzenethiol on Au(111). *J. Phys. Chem. C* **111**, 6335–6342 (2007).
52. Yang, B. *et al.* Chemical Enhancement and Quenching in Single-Molecule Tip-Enhanced Raman Spectroscopy. *Angew. Chem. Int. Ed.* **62**, e202218799 (2023).
53. Liu, S. *et al.* Inelastic Light Scattering in the Vicinity of a Single-Atom Quantum Point Contact in a Plasmonic Picocavity. *ACS Nano* **17**, 10172–10180 (2023).
54. Downes, A., Salter, D. & Elfick, A. Finite Element Simulations of Tip-Enhanced Raman and Fluorescence Spectroscopy. *J. Phys. Chem. B* **110**, 6692–6698 (2006).
55. Chen, X. & Wang, X. Near-field thermal transport in a nanotip under laser irradiation. *Nanotechnology* **22**, 075204 (2011).
56. Zuloaga, J., Prodan, E. & Nordlander, P. Quantum Description of the Plasmon Resonances of a Nanoparticle Dimer. *Nano Lett.* **9**, 887–891 (2009).
57. Marinica, D. C., Kazansky, A. K., Nordlander, P., Aizpurua, J. & Borisov, A. G. Quantum Plasmonics: Nonlinear Effects in the Field Enhancement of a Plasmonic Nanoparticle Dimer. *Nano Lett.* **12**, 1333–1339 (2012).
58. Esteban, R. *et al.* A classical treatment of optical tunneling in plasmonic gaps: extending the quantum corrected model to practical situations. *Faraday Discuss.* **178**, 151–183 (2015).
59. Martín-Jiménez, A. *et al.* Unveiling the radiative local density of optical states of a plasmonic nanocavity by STM. *Nat. Commun.* **11**, 1021 (2020).
60. Savage, K. J. *et al.* Revealing the quantum regime in tunnelling plasmonics. *Nature* **491**, 574–577 (2012).

61. Esteban, R., Borisov, A. G., Nordlander, P. & Aizpurua, J. Bridging quantum and classical plasmonics with a quantum-corrected model. *Nat. Commun.* **3**, 825 (2012).
62. Zhu, W. & Crozier, K. B. Quantum mechanical limit to plasmonic enhancement as observed by surface-enhanced Raman scattering. *Nat. Commun.* **5**, 5228 (2014).
63. Zhu, W. *et al.* Quantum mechanical effects in plasmonic structures with subnanometre gaps. *Nat. Commun.* **7**, 11495 (2016).
64. Jaculbia, R. B. *et al.* Single-molecule resonance Raman effect in a plasmonic nanocavity. *Nat. Nanotechnol.* **15**, 105–110 (2020).
65. De Luca, F. & Ciracì, C. Impact of Surface Charge Depletion on the Free Electron Nonlinear Response of Heavily Doped Semiconductors. *Phys. Rev. Lett.* **129**, 123902 (2022).
66. Wang, Y. *et al.* Direct electrical modulation of second-order optical susceptibility via phase transitions. *Nat. Electron.* **4**, 725–730 (2021).
67. Wang, S. *et al.* Giant electric field-induced second harmonic generation in polar skyrmions. *Nat. Commun.* **15**, 1374 (2024).
68. Caretta, L. *et al.* Non-volatile electric-field control of inversion symmetry. *Nat. Mater.* **22**, 207–215 (2023).
69. Abdelwahab, I. *et al.* Giant second-harmonic generation in ferroelectric NbOI2. *Nat. Photon.* **16**, 644–650 (2022).
70. Takahashi, S. *et al.* Diffraction-Unlimited Tip-Enhanced Sum-Frequency Vibrational Nanoscopy. *arXiv:2509.09179 [physics.optics]*.
71. Seyler, K. L. *et al.* Electrical control of second-harmonic generation in a WSe2 monolayer transistor. *Nat. Nanotechnol.* **10**, 407–411 (2015).
72. Lee, K.-T. *et al.* Electrically Biased Silicon Metasurfaces with Magnetic Mie Resonance for Tunable Harmonic Generation of Light. *ACS Photonics* **6**, 2663–2670 (2019).
73. Gieseking, R. L. M., Lee, J., Tallarida, N., Apkarian, V. A. & Schatz, G. C. Bias-Dependent Chemical Enhancement and Nonclassical Stark Effect in Tip-Enhanced Raman Spectromicroscopy of CO-Terminated Ag Tips. *J. Phys. Chem. Lett.* **9**, 3074–3080 (2018).
74. Lee, J., Tallarida, N., Chen, X., Jensen, L. & Apkarian, V. A. Microscopy with a single-molecule scanning electrometer. *Sci. Adv.* **4**, eaat5472 (2018).
75. Braun, K. *et al.* Probing Bias-Induced Electron Density Shifts in Metal–Molecule Interfaces via Tip-Enhanced Raman Scattering. *J. Am. Chem. Soc.* **143**, 1816–1821 (2021).
76. Chen, S., Li, K. F., Li, G., Cheah, K. W. & Zhang, S. Gigantic electric-field-induced second harmonic generation from an organic conjugated polymer enhanced by a band-edge effect. *Light Sci. Appl.* **8**, 17 (2019).

77. Kalff, F. E. *et al.* A kilobyte rewritable atomic memory. *Nat. Nanotechnol.* **11**, 926–929 (2016).
78. Natterer, F. D. *et al.* Reading and writing single-atom magnets. *Nature* **543**, 226–228 (2017).
79. Xu, X., Gao, C., Emusani, R., Jia, C. & Xiang, D. Toward Practical Single-Molecule/Atom Switches. *Adv. Sci.* **11**, 2400877 (2024).

## Acknowledgements

We are grateful to Prof. A. Morita, Dr. T. Hirano, and Mr. K. Kumagai at the Tohoku University for the fruitful discussions. For technical assistance, we also thank M. Aoyama, T. Kondo, N. Mizutani, T. Kikuchi, and T. Toyoda at the Equipment Development Center, Institute for Molecular Science (IMS), and E. Nakamura at the UVSOR synchrotron facility of IMS. SEM observation of tips was conducted at the Institute for Molecular Science, supported by "Advanced Research Infrastructure for Materials and Nanotechnology in Japan (ARIM)" of the Ministry of Education, Culture, Sports, Science and Technology (MEXT), Proposal Number JPMXP1223MS5022. T.S. acknowledges financial support from JSPS KAKENHI Grant-in-Aid for Scientific Research (A) (19H00865 and 22H00296), JST-PRESTO (JPMJPR1907); JST-CREST (JPMJCR22L2). A.S. acknowledges financial support from JSPS KAKENHI Grant-in-Aid for Scientific Research (B) (23H01855); for Early-Career Scientists (20K15236); Casio Science Promotion Foundation (38-06); and Research Foundation for Opto-Science and Technology. S.T. acknowledges financial support from Grant-in-Aid for JSPS Fellows (22KJ3099).

## Author contributions

S.T. and T.S. conceived the project; S.T., A.S., and T.S. built up the experimental setup; T.M. established the tip fabrication technique; S.T. performed the experiments with help from A.S. and T.M. and analyzed the data; S.T. and T.S. wrote the manuscript; and all authors discussed the results and commented on the manuscript.

### Competing interests

The authors declare no competing financial interest.

## Additional information

**Supplementary information** is available at https://

**Correspondence and requests for materials** should be addressed to A.S. and T.S.

# Supplementary Information:

# Giant near-field nonlinear electrophotonic effects in an angstrom-scale plasmonic junction

Shota Takahashi[1], Atsunori Sakurai[1,2]*, Tatsuto Mochizuki[1,2], and Toshiki Sugimoto[1,2]*

[1] Institute for Molecular Science, National Institutes of Natural Sciences; Okazaki, Aichi 444-8585, Japan.

[2] Graduate Institute for Advanced Studies, SOKENDAI; Okazaki, Aichi 444-8585, Japan.

*Corresponding authors. Email: asakurai@ims.ac.jp; toshiki-sugimoto@ims.ac.jp

## 1. The change in the surface morphology caused by SAM formation.

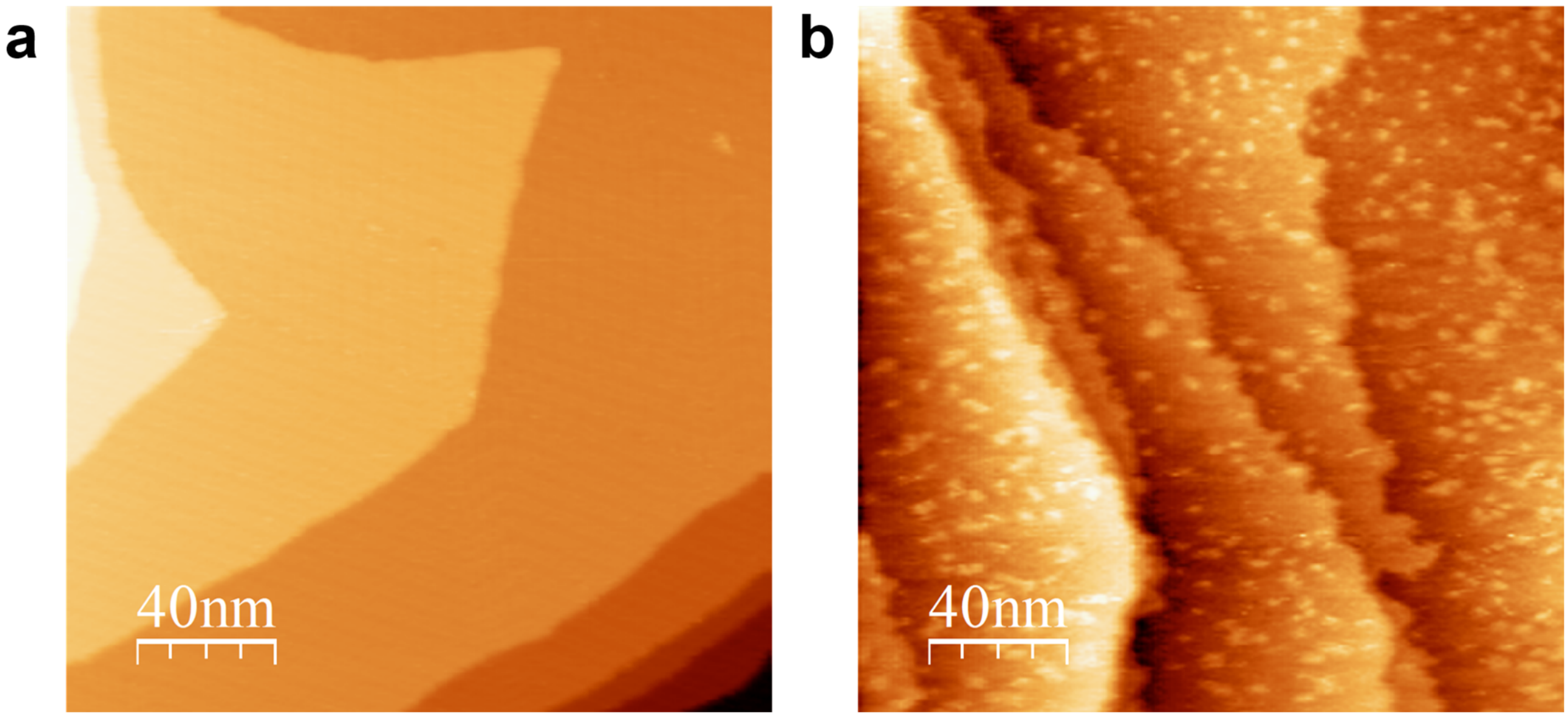

**Fig. S1 | The change in the surface morphology caused by SAM formation.** STM images of Au(111) surface before (**a**) and after (**b**) immersion into ethanolic solution of MBT for 48 hours are shown. These images were obtained using a Au tip in constant current mode under sample bias voltage of –0.5 V. The tunneling current setpoints for (**a**) and (**b**) were 1 nA and 34.6 pA, respectively. The substrate before immersion exhibited the straightened step edges and flat terraces. After immersion, the step edges became jagged, and patch-like protrusions appeared on the terrace. These characteristics are well-known and have also been reported in previous STM studies on aromatic thiolate SAMs[1–3]. Note that the protrusions in **b** were derived from Au adatom islands rather than from molecular aggregates[1–3]. The MBT monolayer is formed not only on the wide terrace region but also on the islands[1–3]. Therefore, the thickness of the SAM layer measured from the topmost gold surface is uniform throughout the surface and the tip-surface distance is always maintained constant during tip scanning at constant tunneling current and voltage.

## 2. Spectra of excitation light

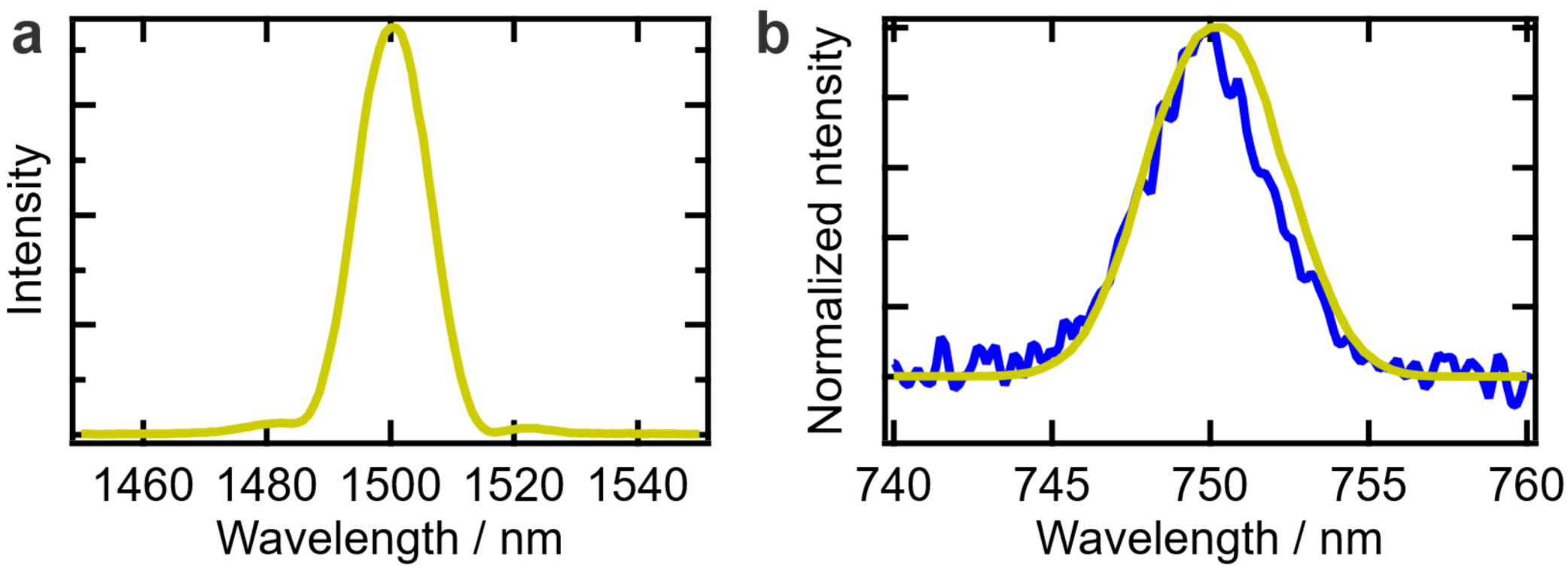

**Fig. S2 | Excitation source for TE-SHG experiments.** (**a**) The spectrum of incident light ("signal" output from optical parametric oscillator (OPO) pumped by 1033-nm fundamental wave) used as excitation source of TE-SHG experiments. The central wavelength is located at 1500 nm. (**b**) Comparison of TE-SHG spectra (dark blue, same curve as shown in Fig. 2a in the main text) and the squared signal output spectrum (dark yellow). In the dark yellow curve, the wavelength axis values were halved to match the spectral positions. The spectral matching indicates that the shape of TE-SHG spectra is determined by the spectral distribution of the excitation light.

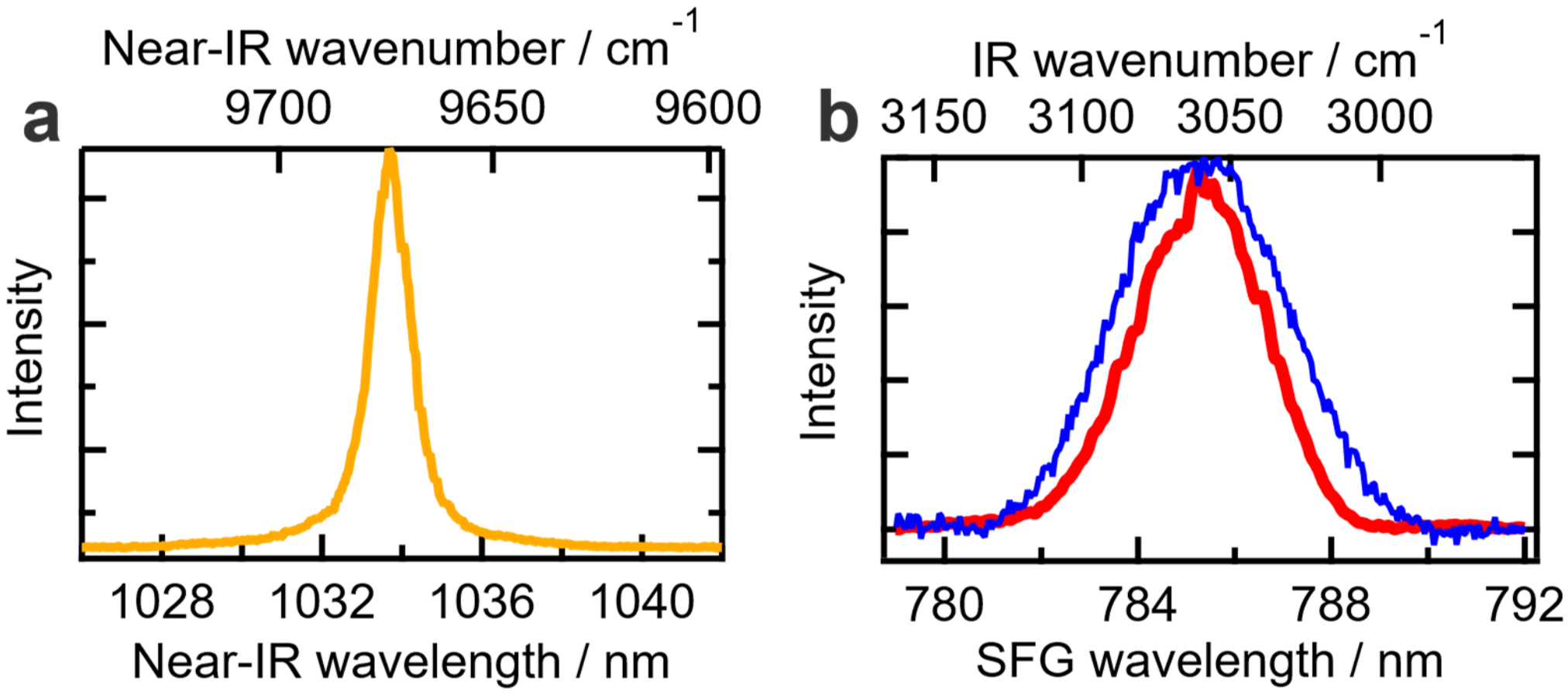

**Fig. S3 | Excitation sources for TE-SFG experiments. a.** The spectrum of near-IR incident light: "fundamental" output from Yb-fiber laser after passing through an air-spaced Fabry-Pérot etalon to narrow down the spectral width. **b.** The spectrum of mid-IR incident light: "idler" output from 1030-nm pump OPO (red). Note that the mid-IR wavenumber is displayed in the top axis, and the corresponding values of the top axis are converted from the mid-IR wavenumber to the SFG wavelength and displayed in the bottom axis. For comparison, TE-SFG spectra (dark blue, same curve as shown in Fig. 4c in the main text) is superimposed. The slightly broadened feature of TE-SFG spectra results from the spectral convolution of the mid-IR pulse with a narrowband 1033-nm pulse shown in **a**. The spectral matching indicates that the shape of TE-SFG spectra is determined by the spectral distribution of these two excitation lights.

**3. Estimation of the absolute tip–substrate gap distance.**

The absolute gap distance between the tip apex and metal surface ($d$) is an important experimental parameter for tip-enhanced nonlinear optical measurements. This gap distance was estimated by the $I_t - d$ curve (Fig. S4) for the Au tip and the Au substrate covered with SAM, which was measured by monitoring the tunneling current value ($I_t$) with moving the tip from the initial position ($\Delta d = 0$) toward the substrate under constant sample bias ($V = 0.1$ V). In the present study, the position of $\Delta d = 0$ is defined by the condition of tunneling current setpoint ($I_t$) of 10 pA and sample bias ($V$) of 0.1 V. Although the initial stage of the approach curve ($|\Delta d| < 1.5$ Å) can be well fit with an exponential function, the approach curve exhibits deviation from the exponential behavior after reaching to $\Delta d = -1.5$ Å that gives $I_t \sim 0.12$ nA. Similar results were reported and reasonably explained considering that the vacuum layer and the SAM layer have independent decay constants for the electron tunneling[4–6]. Therefore, the contact point of the tip apex and the SAM should be the point at $\Delta d = -1.5$ Å. Considering the thickness of the MBT molecular layer (~6 Å), the absolute distance between the Au tip and the Au substrate surface at this contact point ($\Delta d = -1.5$ Å) can be determined as approximately 6 Å.

Based on this estimation, we can determine the absolute gap distance corresponding to the TE-SHG spectrum measured at $V = 0.1$ V and $I_t = 0.5$ nA (light blue curve in Fig. 2a in the main text). According to Fig. S4, the tip position defined by $V = 0.1$ V and $I_t = 0.5$ nA corresponds to the point at $\Delta d = -2.7$ Å, which is 1.2 Å below the contact point. In this case, the absolute distance between the Au tip and Au substrate surface can be estimated as $6\text{ Å} - 1.2\text{ Å} = 4.8\text{ Å} \sim 5\text{ Å}$, with the tip apex slightly penetrating into the SAM. It should be noted that such direct contact conditions can sometimes give rise to additional signal enhancement effects arising from chemical bonding formation between the tip and sample[7]. However, since the terminal methyl group of MBT molecules is chemically inert, and its interactions with the tip are too weak to form chemical bonds, such additional enhancement should be negligibly small in our measurement conditions. Indeed, even when the metal–to–metal distance reached ~5 Å where the tip–molecule contact was expected, we did not observe any significant increase in signal intensity (the light blue curve in Fig. 2a in the main text). Therefore, the contribution of contact-induced chemical enhancement can be disregarded in this study.

Then, when the sample bias was increased from 0.1 V to 0.75 V under a constant tunneling current mode (0.5 nA), the tip–substrate gap distance was elongated by 2.1 Å (Fig. 2b in the main text). Therefore, the absolute tip–substrate gap distance at $V = 0.75$ V and $I_t = 0.5$ nA (dark blue curve in Fig. 2a in the main text) can be estimated as $4.8\text{ Å} + 2.1\text{ Å} = 6.9\text{ Å} \sim 7\text{ Å}$.

The absolute gap distances corresponding to the bias-dependent TE-SFG spectra measured under the constant tunneling current of 0.25 nA (Fig. 4c in the main text) can also be estimated in a similar way. According to Fig. S4, the tip position defined by $V = 0.1$ V and $I_t = 0.25$ nA corresponds to the point at $\Delta d = -2.0$ Å, which is 0.5 Å below the contact point. In this case, the absolute distance between the Au tip and Au substrate surface can be estimated as $6\text{ Å} - 0.5\text{ Å} = 5.5\text{ Å}$. Then, when the sample bias was increased from 0.1 V to 0.25 V under a constant tunneling current mode (0.25 nA), the tip–substrate gap distance was elongated by 1.2 Å (Fig. S5). Therefore, the absolute tip–substrate gap distance at $V = 0.25\ \mathrm{V}$ and $I_t = 0.25\ \mathrm{nA}$ (cyan curve in Fig. 4c in the main text) can be estimated as $5.5\ \mathrm{Å} + 1.2\ \mathrm{Å} = 6.7\ \mathrm{Å}$. Similarly, according to the $d - V$ curve in Fig. S5, the distances at $V$ = 0.5 V (sky blue curve in Fig. 4c) and $V$ = 0.75 V (dark blue curve in Fig. 4c) under $I_t = 0.25\ \mathrm{nA}$ are estimated as 6.7 Å + 0.9 Å = 7.6 Å and 7.6 Å + 0.6 Å = 8.2 Å, respectively.

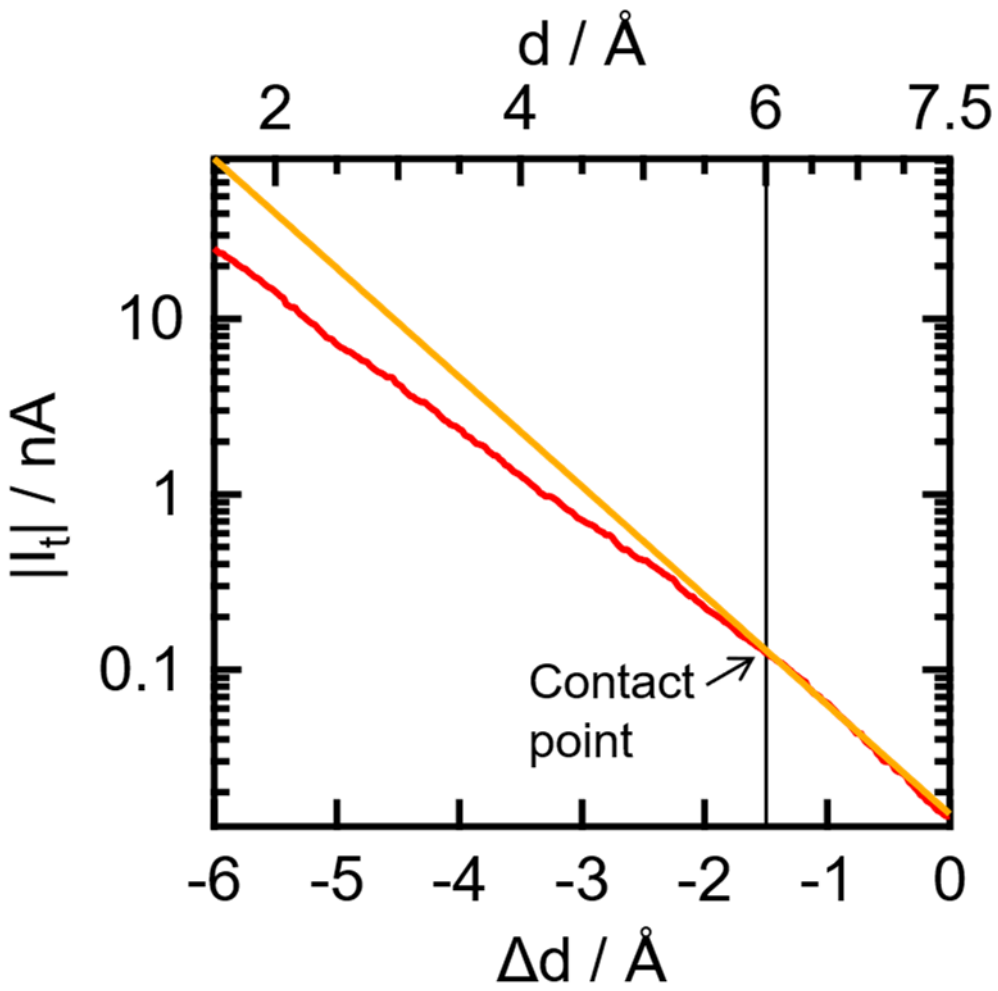

**Fig. S4 | $I_t - d$ curve for the tip–substrate gap.** Variation of the tunneling current ($I_t$) with tip movement ($I_t - d$ curve) were obtained at the sample bias ($V$) 0.1 V. $I_t$ is shown on a logarithmic scale. The initial position ($\Delta d = 0$) is defined by the conditions of $I_t = 10\ \mathrm{pA}$ and $V = 0.1\ \mathrm{V}$. The absolute tip–substrate distance $d$ is displayed in the top axis. The orange line is the fitting curve with an exponential function for the initial stage of the experimental $I_t - d$ curve.

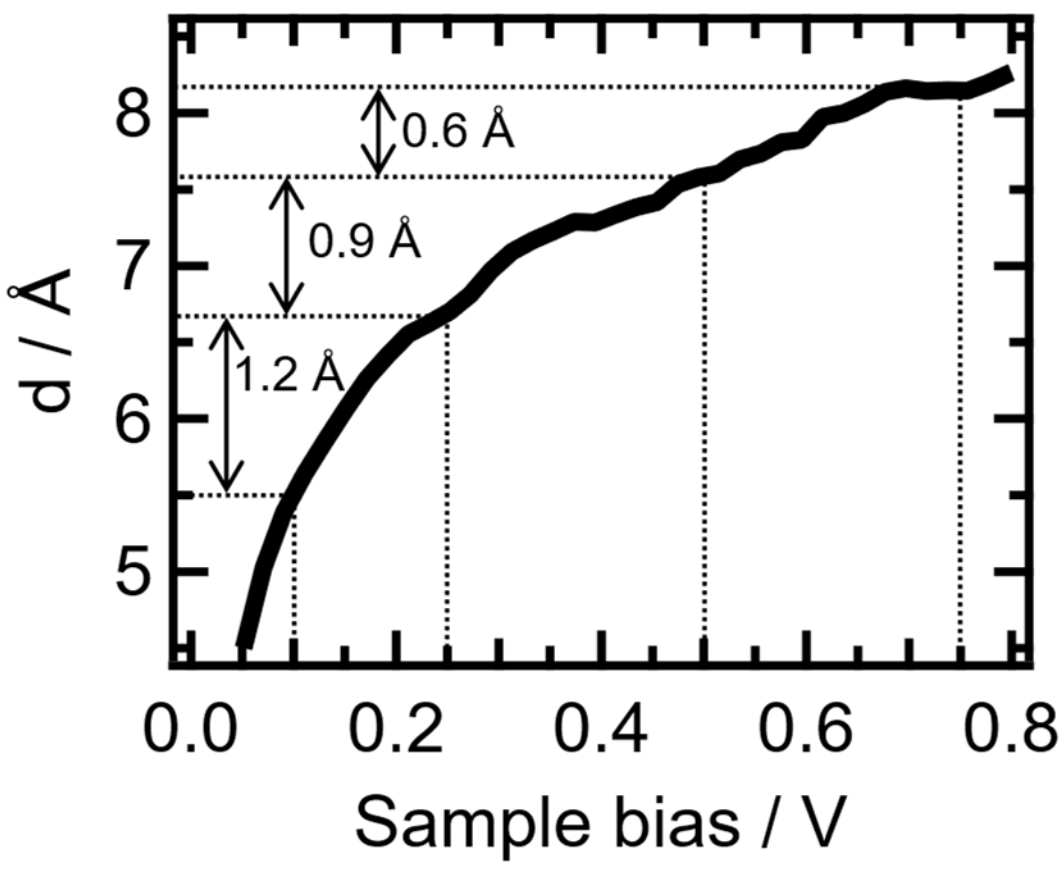

**Fig. S5 | Bias-induced change in the tip-sample distance.** The absolute tip-sample distance ($d$) obtained at a constant tunneling current of 250 pA is plotted as a function of the sample bias. The tip–substrate distance at the feedback loop parameters of $I_t$ = 250 pA and $V$ = 0.1 V is 5.5 Å. The bias increase elongates the tip–substrate distance. Specifically, the distances at 0.25 V, 0.5 V, and 0.75 V are 6.7 Å, 7.6 Å, and 8.2 Å, respectively, which correspond to the experimental conditions where the TE-SFG spectra in Fig. 4c in the main text were obtained.

## 4. Fixing the tip–substrate distance for the measurements of voltage dependences

In the main text, we investigated the dependence of TE-SHG and TE-SFG intensities on applied voltages. In these experiments, it was critical to minimize the voltage-dependent changes in the gap distance between the tip apex and metal surface ($d$) to reduce the variation of field enhancement strength and isolate purely voltage-induced effects. The relationship between tip–substrate distance $d$, sample bias $V$, and tunneling current $I_t$ can be described based on the Landauer theory of the conductance quantum as follows[8]:

$$\frac{I_t}{V} = G_0 \exp(-2\kappa d), \tag{S1}$$

where $G_0$ and $\kappa$ represent the quantum conductance and tunneling decay constant, respectively. Equation (S1) indicates that keeping tip–substrate distance ($d$) across various STM voltages ($V$) requires the tuning of the tunneling current setpoint ($I_t$) in synchronization with STM voltage sweeps. Therefore, prior to the voltage-dependent TE-SHG/TE-SFG measurements, we identified the combinations of $V$ and $|I_t|$ values that can preserve gap distances using the following procedure.

Figure S6a shows typical STM $d - V$ curves obtained for MBT SAM on Au substrate, with two distinct $|I_t|$ values (60 pA and 600 pA). We here consider a horizontal line at $d$ = 7 Å, the distance where the tip is placed 1 Å above the MBT molecules. The intersection points of this line with the $d - V$ curves correspond to the combination of $V$ and $|I_t|$ setpoints that can maintain the tip–substrate distance at $d$~7 Å. For example, in Fig. S6a, four such combinations are derived (orange cross marks): (–0.754 V, 600 pA), (–0.142 V, 60 pA), (+0.147 V, 60 pA), (+0.755 V, 600 pA). Using these combinations, the tip–surface distance $d$ can be effectively kept constant although the voltage is varied.

By repeatedly measuring $d - V$ curves for various $|I_t|$ setpoints, we determined multiple combinations of $V$ and $|I_t|$ values by identifying their intersections with the horizontal line at $d$ = 7 Å (Fig. S6b). Plotting these combinations produces the calibration curve ($I_t - V$) along which $d$ is kept at 7 Å (Fig. S6c). The voltage-dependent TE-SHG experiments shown in Fig. 3 in the main text were performed by selecting $V$ and $|I_t|$ values along this calibration curve. The explicit $V$ and $|I_t|$ values employed in our TE-SHG measurements are listed in Table S1.

Additionally, as shown in Fig. S6d and e, a similar calibration was performed again prior to the voltage-dependent TE-SFG experiments (Fig. 4d and e in the main text). The difference of $d - V$ curves shown in Fig. S6b and e and the resultant calibration curve (Fig. S6c and S5e) can be attributed to the variation in the tip apex structures: the difference in the tips influenced the $d - V$ characteristics of the gap and altered the appropriate combination of $V$ and $|I_t|$ required to maintain a constant $d$. The explicit $V$ and $|I_t|$ values adopted in our TE-SFG measurements are also listed in Table S1.

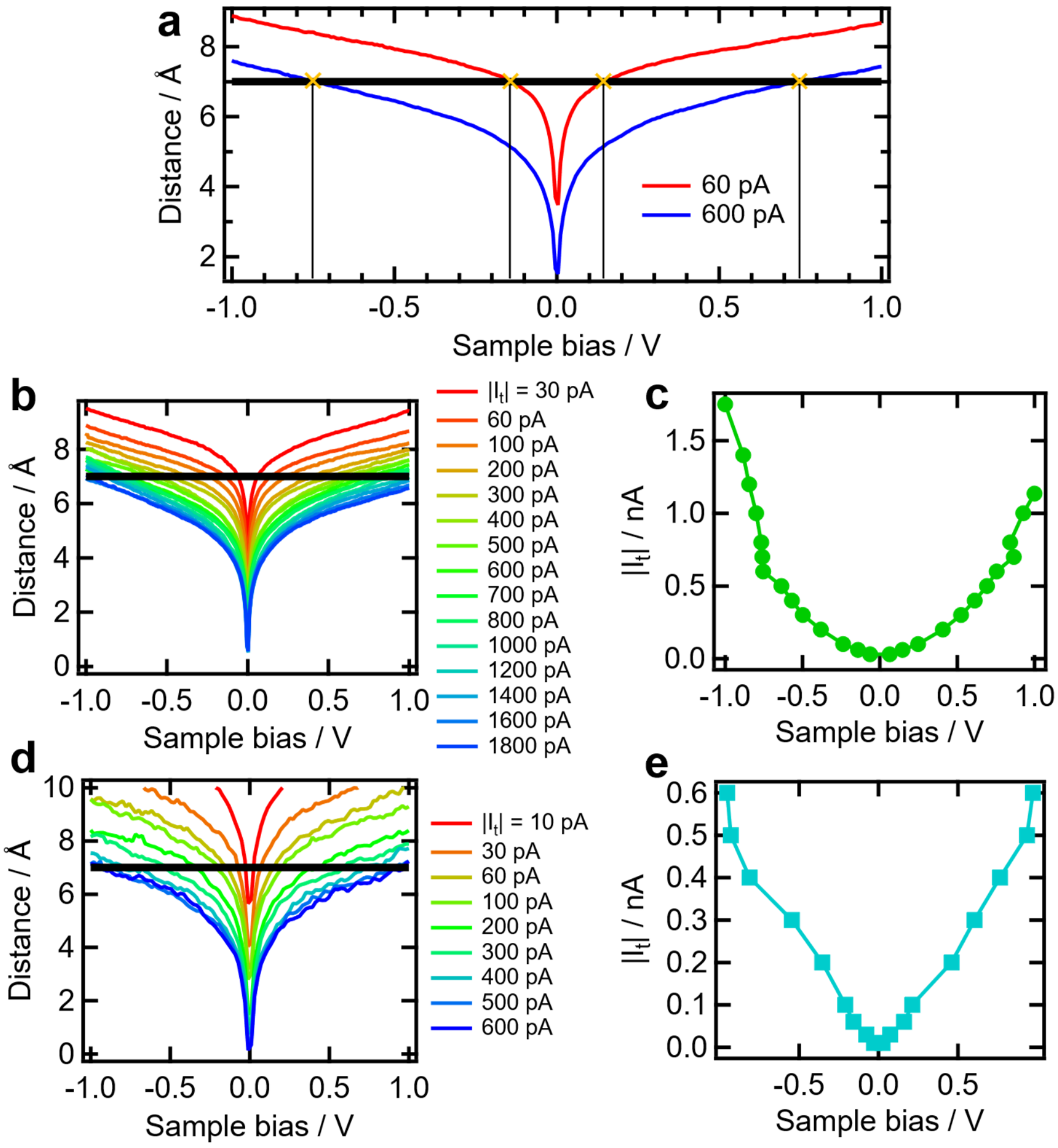

**Fig. S6 | Deducing $V$ and $|I_t|$ values to fix the tip–surface distance. a**. Typical STM $d-V$ curves obtained for MBT SAM on Au substrate by using two different $|I_t|$ values (60 pA and 600 pA). The black horizontal line indicates the tip–surface distance of 7 Å. The orange cross marks represent the intersections between the line of $d = 7$ Å and $d-V$ curves. The combinations of $V$ and $|I_t|$ setpoints at these intersections allow us to maintain $d$~7 Å. **b, d**. $d-V$ curves for various $|I_t|$ setpoints. The black horizontal line indicates the tip–surface distance of 7 Å. The data in **b** and **d** were obtained with the tips used for voltage-dependent TE-SHG and TE-SFG experiments, respectively. **c, e**. The $I_t-V$ calibration curve obtained by tracking the intersection points of $d =$ 7 Å line with the $d-V$ curves. The voltage-dependent TE-SHG and TE-SFG measurements shown in the main text were performed by selecting the pairs of $V$ and $|I_t|$ values along the curves in **c** and **e**, respectively.

**Table S1** Tunneling current setpoints at different sample biases to fix the tip–substrate distance at $d$~7 Å

| Bias | -1 V | -0.75 V | -0.5 V | -0.25 V | -0.1 V | 0.1 V | 0.25 V | 0.5 V | 0.75 V | 1 V |
|---|---|---|---|---|---|---|---|---|---|---|
| TE-SHG | 1750 pA | 595 pA | 304 pA | 111 pA | 47.2 pA | 44.6 pA | 102 pA | 277 pA | 591 pA | 1135 pA |
| TE-SFG | - | 381 pA | 254 pA | 127 pA | 50.8 pA | 50.8 pA | 127 pA | 254 pA | 381 pA | - |

## 5. Forward-scattered TE-SHG signals

Here, we briefly discuss the characteristic features of the forward-scattered TE-SHG signals by comparing them with the backward-scattered signals shown in the main text. Figure S7 shows the spectra obtained in the forward-scattering geometry under the irradiation of 10 pJ excitation light, which were measured simultaneously with the data shown in Fig 2a in the main text. When the tip–substrate distance was approximately 30 nm, the coherent far-field SHG signal was observed in the forward-scattering direction (gray curve in Fig. S7), whereas no appreciable backward-scattered signal was observed (gray curve in Fig. 2a in the main text). These results allow us to exclude the possibility of the dominant effect of the tip plasmons excited at the Au tip apex alone, because if the tip plasmons could facilitate the appreciable level of enhanced SHG, the signals should be observed both in the forward- and backward-scattering geometry due to the dipole-like pattern of the plasmonic radiation. Therefore, the exclusive observation of forward-scattered SHG clearly indicates that the contribution from the tip plasmons excited at the Au tip apex alone is negligibly small and below the detection threshold. Such negligibly small contribution from the tip plasmon is more quantitatively supported by our theoretical calculation in Section 16, which reveals that the field enhancement strength and radiation efficiency under the 30-nm tip–substrate distance are more than one order of magnitude weaker than those of angstrom-scale junction plasmon (see Section 16 for details).

Then, the SHG intensity increased when the tip-substrate distance was reduced from 30 nm to ~5 Å (light blue curves in Fig. S7 and Fig. 2a in the main text). As discussed in the main text, this is the manifestation of optical enhancement effect caused by the formation of plasmonic nanocavity between the substrate and the apex of the tip. Moreover, similarly to the backward-scattered signals shown in Fig. 2a in the main text, the forward-scattered TE-SHG also further increased when the sample bias ($V$) was increased from 0.1 V to 0.75 V (dark blue curve). Therefore, except for the existence of normal far-field signal, forward-scattered TE-SHG signals also exhibit similar enhancement behavior and voltage response.

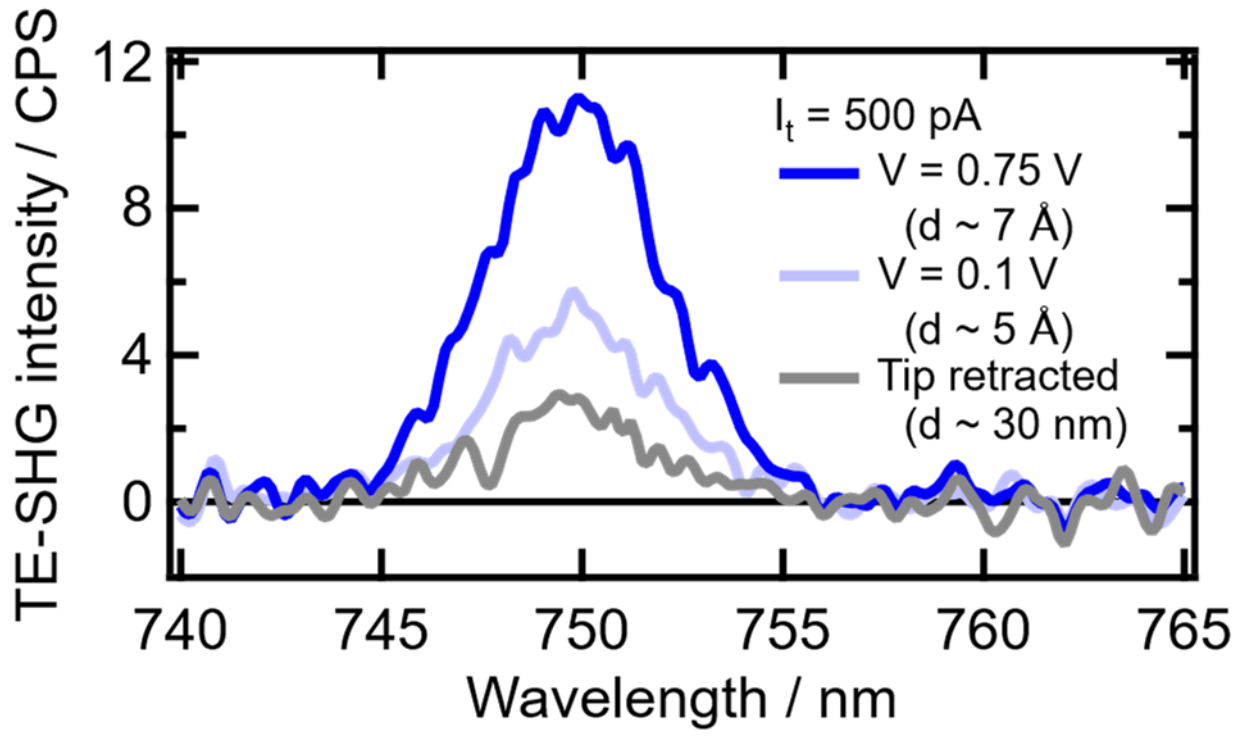

**Fig. S7 | The spectra of forward-scattered TE-SHG.** These data were obtained at 1500-nm excitation and at sample biases of 0.1 V (light blue, $d$~5 Å) and 0.75 V (dark blue, $d$~7 Å) with a constant tunnelling current of 500 pA under room temperature and ultra-high vacuum (<1×10$^{-7}$ Pa) condition. Gray curve indicates the signal obtained when the substrate was retracted enough from the tip ($d$~30 nm) to deactivate the plasmonic enhancement effects.

**6. The stability of tip-enhanced nonlinear optical signals**

In this section, we discuss the stability of tip-enhanced nonlinear optical signals by examining the individual results of repeated measurements. The voltage dependence of TE-SHG intensities shown in Fig. 3 in the main text was obtained by multiple measurements repeated three times. The results of those three measurements are individually presented in Fig. S8. Importantly, the different voltage sweeping orders were employed in each measurement. In the first measurement, the voltage was swept sequentially from −1 V to +1 V (filled orange squares). In the second measurement, the voltage was changed in the reverse direction, from +1 V to −1 V (open blue diamonds). In the third measurement, the voltage was initially swept from −0.1 V to −1 V, followed by a switch to +1 V and then sequentially reduced to +0.1 V (open green triangles). Despite these varied voltage sweeping sequences, the TE-SHG signals consistently exhibited a quadratic dependence on the applied voltage. This clearly indicates the minimal optical damaging effects on the tip during the measurements and ensures the reliability of our experimental results. Note that the error bars shown in Fig. 3 in the main text were obtained through these three measurements.

The stability of TE-SFG signals was also ensured in a similar way. The voltage dependence of TE-SFG intensities shown in Fig. 4d and e in the main text was obtained by sweeping the voltage in a random order. The explicit sequence of the voltage sweeping is displayed in Table S2. Despite such random-order voltage sweeping, the TE-SFG signals exhibited clear quadratic dependence on the applied voltage. Note that under the individual bias voltages, 30-second signal accumulation was repeated ten times. The error bars shown in Fig. 4e in the main text were obtained through these ten measurements. More importantly, after performing measurements for eight different voltage values, we switched the voltage to the initial value (+0.75 V) again and confirmed that the TE-SFG intensity was reproduced (Fig. S9). It should also be noted that this verification of the signal stability was performed under air atmosphere at room temperature. On the basis of the high reproducibility under the ambient conditions, we can conclude that the quadratic behavior is not due to accidental temporal change but is the manifestation of the intrinsic field-effect modulation of angstrom-scale gap structures that is operable even under the ambient conditions.

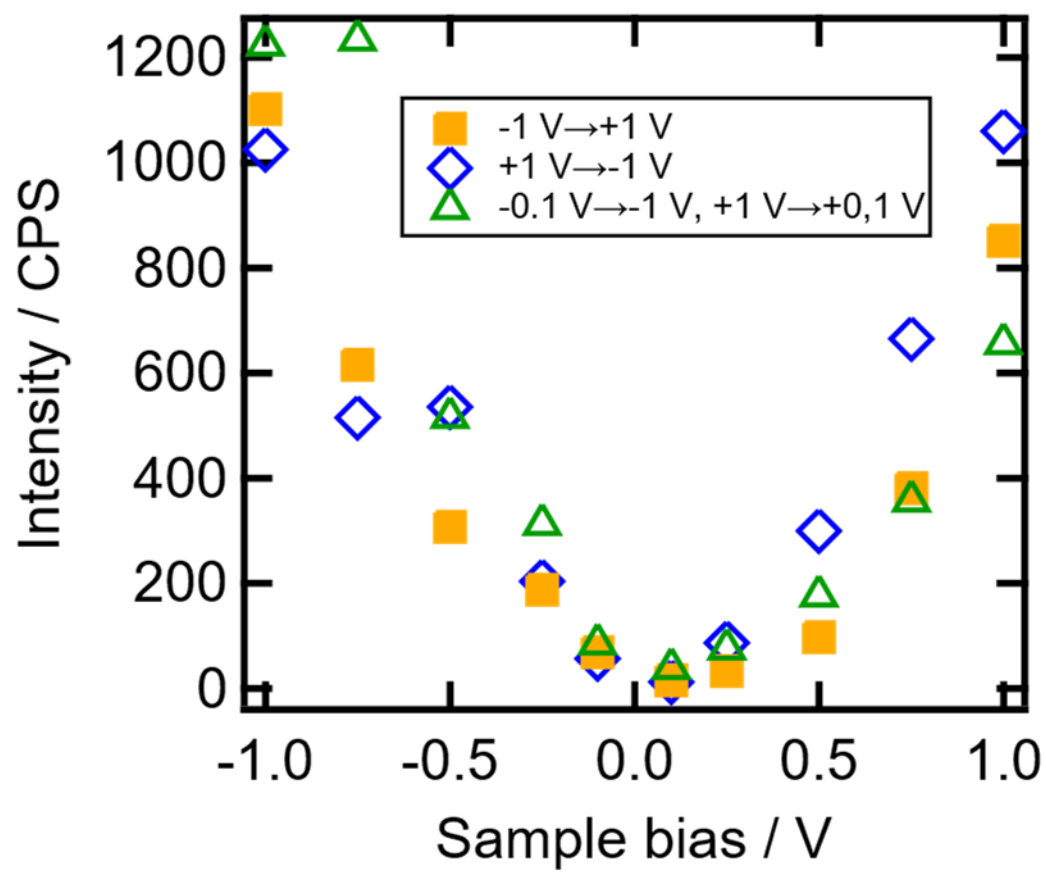

**Fig. S8 | Stability of TE-SHG measurements.** The voltage-dependent changes in the TE-SHG intensity observed for three different voltage sweeping schemes are shown.

**Table S2** Voltage sweeping order employed in the voltage dependence measurements of TE-SFG.

| Measurement order | 1 | 2 | 3 | 4 | 5 | 6 | 7 | 8 | 9 |
|---|---|---|---|---|---|---|---|---|---|
| Voltage / V | +0.75 | −0.10 | +0.25 | +0.50 | −0.25 | −0.75 | −0.50 | +0.10 | +0.75 |

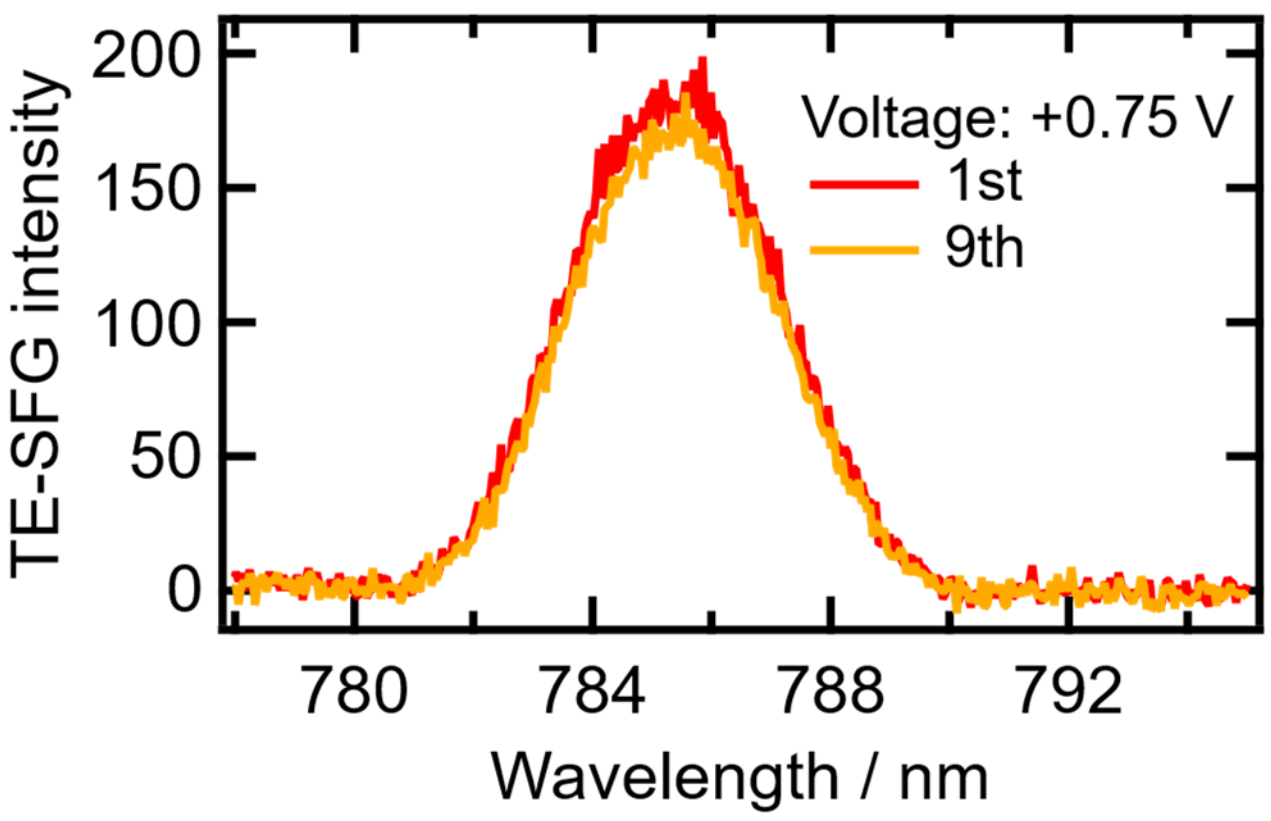

**Fig. S9 | Stability of TE-SFG measurements.** TE-SFG spectra obtained under +0.75 V sample bias voltage. The measurements were performed in ambient condition. The red curve represents the initially obtained result (measurement order 1 in Table S2). The orange curve represents the result finally obtained after performing measurements for eight different voltage values (measurement order 9 in Table S2). Note that the TE-SFG spectrum at +0.75 V shown in Fig. 4d in the main text was given by the average of these two spectra.

## 7. Conversion efficiencies of TE-SHG and TE-SFG responses

In this section, we discuss the conversion efficiencies of TE-SHG and TE-SFG responses of the SAM-adsorbed gold substrate (Fig. S10). Based on the FF-SHG intensity spectrum shown in Fig. S7, the conversion efficiency of FF-SHG is estimated to be on the order of $10^{-14}$. In contrast, the TE-SHG and TE-SFG exhibited higher efficiencies on the order of $10^{-11}$ and $10^{-12}$, respectively (Fig. S10), indicating that the nonlinear optical efficiency is enhanced by near-field electric field confinement. Importantly, this enhancement occurs despite the extremely small number of molecules involved in the TE-SHG/TE-SFG processes: based on the ~10-nm-scale field enhancement area (Fig. S19) and the previously reported molecular density of the MBT SAM ($\sim 4 \times 10^{14}$ $cm^{-2}$)[9], only $\sim 10^3$ molecules are estimated to contribute to TE-SHG/TE-SFG process. In contrast, the FF-SHG process should involve at least $\sim 10^9$ molecules within the micrometer-scale optical focus spot. Thus, while the absolute efficiencies are low, the per-molecule efficiency in the near-field scheme ($10^{-14}$–$10^{-15}$) is significantly higher than that in the far-field condition ($\sim 10^{-23}$), clearly demonstrating the crucial role of the near-field enhancement ($10^9$–$10^8$) in the nonlinear optical generation processes. It should be noted that $\sim 10^3$ molecules are still too few to satisfy macroscopic phase-matching conditions, and the TE-SHG/TE-SFG signals under the present experimental conditions can therefore be approximated as originating from single-dipole radiation[10].

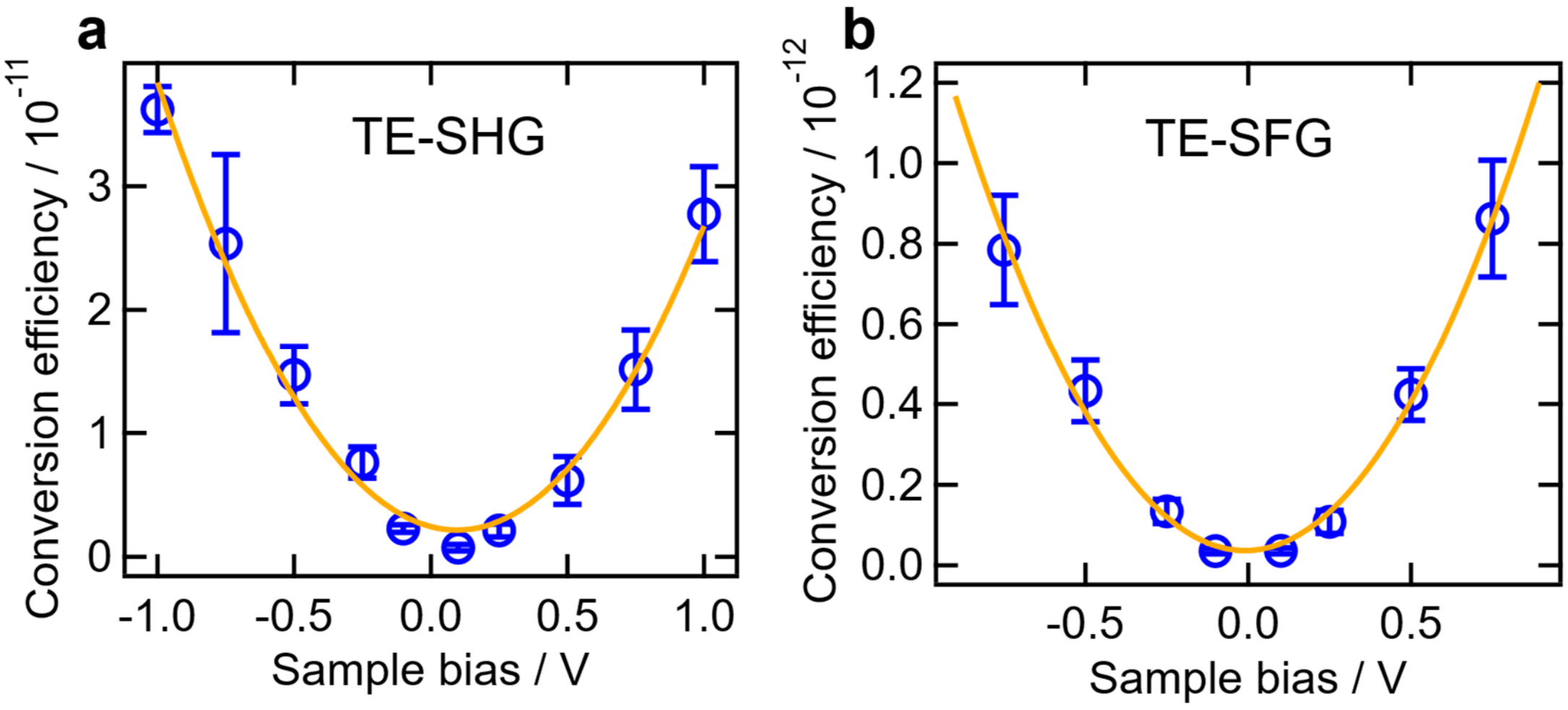

**Fig. S10 | Conversion efficiencies of TE-SHG (a) and TE-SFG (b) processes plotted against the sample bias.** The conversion efficiency data presented in panel (**a**) were obtained by converting the photon counts used to evaluate the modulation depth in Fig. 3 in the main text into optical power and normalizing them by the incident power. The data in panel (**b**) were calculated in the same manner from the measurements shown in Fig. 4e in the main text.

**8. STM images with and without laser irradiation—Negligibly small optical damage of the tip produced by excitation laser—**

To ensure that the exposure to excitation laser does not affect or damage the structure of the tip apex, we compared STM images of MBT-adsorbed Au(111) surface with and without 0.5 mW near-IR pulse irradiation (Fig. S11). Although the STM images shown in Fig. S11a and b were obtained at different locations on the substrate, the noise levels on both images were sufficiently low and angstrom-scale Au monoatomic step structures whose heights are consistent with previously reported values[11–13] were clearly captured (Fig. S11c and d). Therefore, the fact that we were able to obtain such stable and high-quality STM images under laser illumination strongly supports the conclusion that the tip apex remains structurally stable during the measurements.

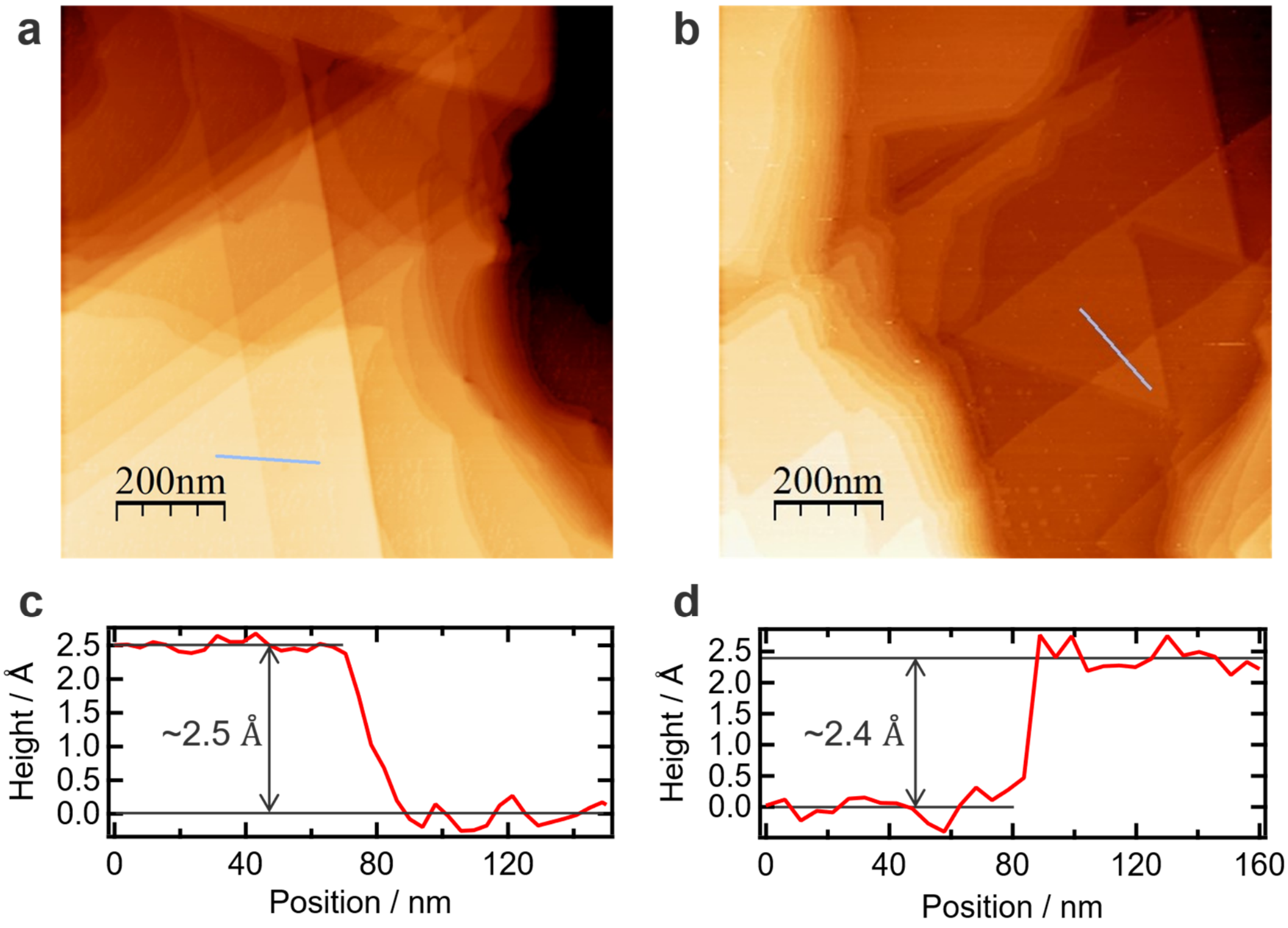

**Fig. S11 | The effect of laser irradiation on STM measurements.** STM images of Au(111) with the adsorption of MBT SAM obtained without (**a**) and with (**b**) irradiating the tip-substrate gap region by near-IR excitation laser for TE-SHG experiments (50 MHz, 1500 nm, 0.5 mW). These STM images were measured at different locations on the substrate. The combinations of the bias voltages and tunneling current set points for (**a**) and (**b**) were (0.1 V, 1 nA) and (1 V, 0.2 nA), respectively. Both images clearly capture single-atom step structures with similar quality and low noise levels, confirming that the optical damage of the tip produced by excitation laser is negligibly small. **c** and **d**. Height profiles along the light blue lines in **a** and **b**, respectively.

## 9. The absence of bias-induced structural change in MBT SAM

Previous experimental work demonstrated that MBT SAM on Au(111) substrate immersed in electrolyte solution exhibits drastic STM bias-induced structural transformation at approximately +0.3 V, accompanied by significant changes in the number and density of island-like structures within several minutes[9]. In contrast, in our ultra-high vacuum and air experimental conditions, such drastic structural change was not observed even when applying +1 V STM bias for more than 10 minutes (Fig. S12). Therefore, we can reasonably assume that the molecular-level structure of the MBT SAM and its intrinsic second-order nonlinear optical susceptibility remain unchanged throughout our experiments.

The voltage-dependent TE-SHG/SFG measurements (Fig. 3 and 4 in the main text) were performed under these stable conditions, where no temporal structural changes in the SAM film were detected. Importantly, in these measurements, the bias values were varied in random orders (see Supplementary Section 6). Despite such non-sequential bias sweeping, the modulation curves exhibited a quadratic dependence on the applied voltage (Fig. 3 and 4 in the main text), and those experimental data were well-fit by using specific $\chi^{(2)}$ and $\chi^{(3)}$ values. This guarantees that no structural modifications occurred in the SAM during the measurements of voltage-dependent TE-SHG/SFG measurements, where the voltage values were swept within the range of –1 V to +1 V.

Notably, the absence of the structural modification of the tip and substrate indicates that dielectric breakdown within the gap and associated breakdown-induced damage to either the sample or the tip can be ruled out, even under the application of a strong electrostatic field (~$10^9$ V/m) across the gap. The impossibility of the dielectric breakdown can be correctly interpreted by considering the probability of the electron–gas collision within our angstrom-scale gap. Generally, the dielectric breakdown is induced by field emission of electrons from an electrode into gas phase, avalanche ionization of gas molecules, and ballistic acceleration of charge carriers within the interelectrode gap, leading to physical damage of the electrodes[14]. Thus, the collision process between electrons and gas molecules plays a key role in the emergence of dielectric breakdown. However, since the angstrom-scale tip–substrate distance employed in our experiments ensures the overlap of electronic wavefunctions at the tip and substrate surfaces, the interelectrode electron transfer occurs predominantly via the tunneling process, rather than ballistic field emission of electrons into the gas phase. This prevents electron–gas collisions that would otherwise trigger avalanche ionization of gas molecules, resulting in the absence of the dielectric breakdown. Moreover, considering the number density of gas molecules under ambient conditions ($2.69\times10^{19}$ cm$^{-3}$), the extremely small volume of the angstrom-scale gap (~$10^{-20}$ cm$^3$) allows for the presence of at most a few gas molecules within the gap. This is insufficient to sustain the cascade ionization processes, making gas-phase electric breakdown physically impossible even under ambient experimental conditions. Therefore, the possibility of electric discharges within our angstrom-scale gap structure can be clearly ruled out, and any discharge-induced modification of the tip or substrate can be safely disregarded.

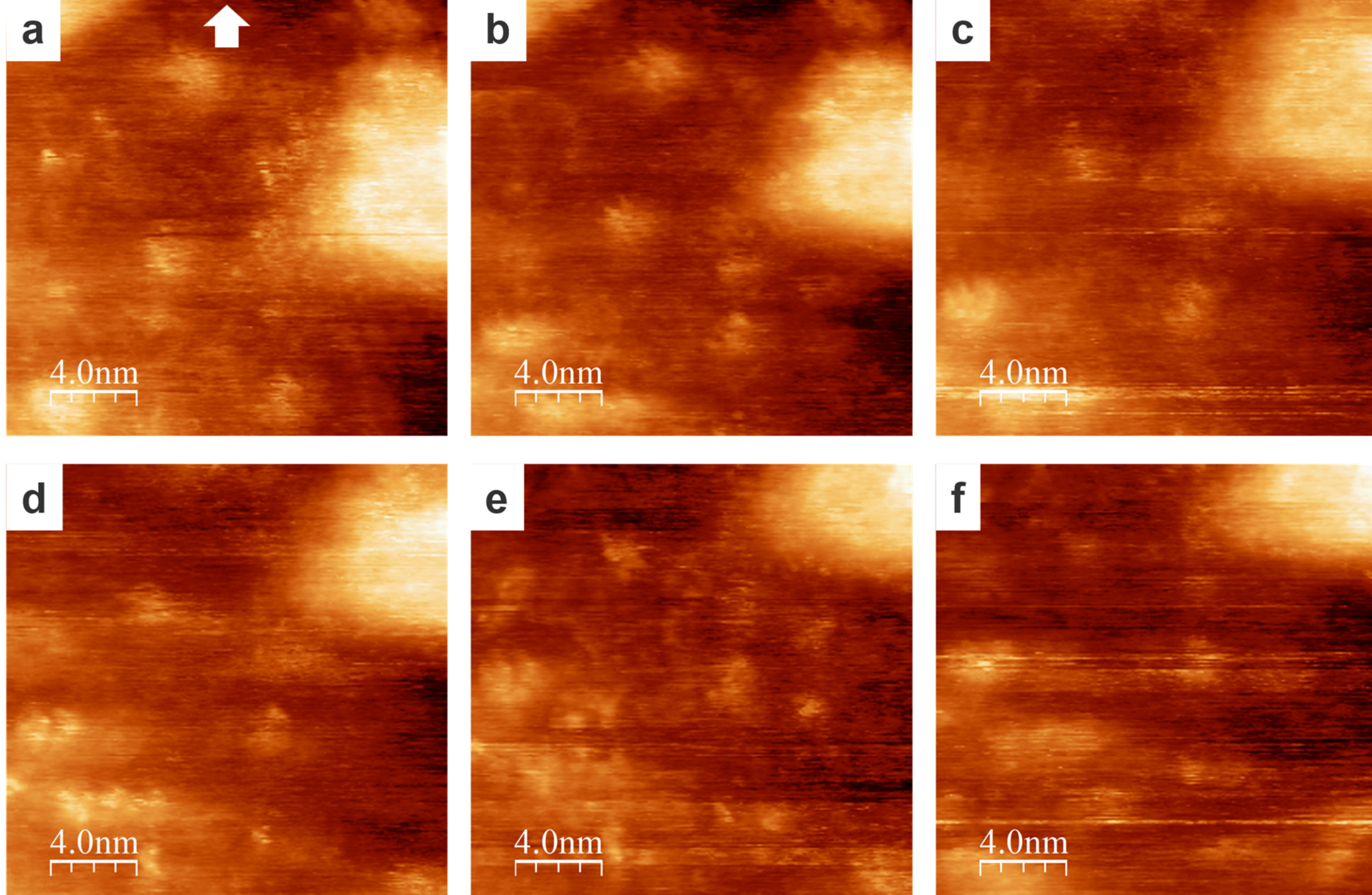

**Fig. S12 | Time-lapse STM images of MBT SAM on Au(111).** The images were obtained sequentially from **a** to **f** under the sample bias of +1 V and tunneling current setpoint of 10 pA. The island-like structures that can be seen in each image correspond to the magnified view of the patch-like protrusions shown in Fig. S1b. These structures were derived from Au adatom islands rather than from bumps of molecular aggregates, and the MBT monolayer is also formed on these islands[1–3]. The tip-substrate gap was not irradiated by excitation lasers. Since the acquisition time for each image was 105 seconds, it took more than 10 minutes to obtain all images. The thermal drift of piezoelectric stage where the sample was mounted gradually shifts the imaging region. The white arrow in **a** indicates the direction of the thermal drift of the piezoelectric stage.

## 10. Origin of near-field nonlinear optical signals

To identify the origin of the near-field nonlinear optical signals presented in the main text, we conducted additional TE-SHG experiments using a bare Au(111) surface without adsorption of SAM molecules. As shown in Fig. S13, the bare Au(111) surface also generated TE-SHG signals and exhibited large modulation depth on the order of 1000%/V, which is comparable to the case of the SAM-embedded gap. This finding indicates that the plasmonic gap structure itself plays a crucial role in the generation of tip-enhanced nonlinear optical signals and their field-induced modulation. We currently consider that the TE-SHG signals observed in the absence of the SAM film arise from the surface electrons of the Au tip and substrate, seeping out ∼2 Å from their surfaces.

It is also noteworthy that the absolute TE-SHG signal intensity for the Au(111) substrate coated by SAM molecules was much higher than that of the bare Au(111) substrate (Fig. S13a). Based on this result, we consider that the SAM film mainly contributes to the SHG signal generation. The difference in the absolute TE-SHG intensity for the Au substrates with and without the SAM can be understood from the spatial distribution of the electric field within the gap. In a previous study, electric field distribution within the gap was precisely calculated by incorporating atomic arrangements and charge distributions in an angstrom-scale gaps, predicting that the field enhancement at the interface between metal surface and vacuum region is generally weaker than that at the center of the gap region[15]. Thus, the SAM film within the gap should experience a substantially stronger electric field compared to the surface electrons. This pronounced difference in the strength of the electric field experienced by the media is likely a primary factor contributing to the variation in TE-SHG intensity shown in Fig. S13a.

Additionally, the observed differences in TE-SHG intensity (Fig. S13a) may also be caused by the differences in the effective nonlinear susceptibilities of the SAM film and the surface electron system. Moreover, in the case of bare Au (111) substrate, the electron wavefunctions spilling out from the tip apex and substrate surface are expected to more largely overlap compared with the SAM-covered configuration. Such overlap would enhance quantum plasmonic quenching effects[15–21], leading to a weaker near-field enhancement and consequently a smaller TE-SHG signal for the bare Au(111) substrate compared with the SAM-covered surface. A more quantitative analysis of these effects is beyond the scope of the present study and will be addressed in future work. Nevertheless, considering these possible factors, using the SAM film as the optical medium and positioning it within the gap region appears to be highly advantageous for enhancing TE-SHG signal.

Based on the above discussion, the overall mechanism governing the observed TE-SHG/TE-SFG and their bias-induced modulation can be comprehensively described as follows. While the nonlinear susceptibilities $\chi^{(2)}$ and $\chi^{(3)}$ originate from the medium within the gap (either the SAM or surface electron systems), plasmonic field enhancement remains indispensable for amplifying and detecting the near-field optical responses from the angstrom-scale gap region. Crucially, the exceptionally high bias-induced modulation depth (~2000%/V) observed in our study is fundamentally driven by the interaction between the third-order $\chi^{(3)}$ of the medium and the intense electrostatic field ($E_{\mathrm{DC}}$) induced through voltage application. Although the absolute intensity of near-field nonlinear optical signals is determined by a complex interplay of various factors, including the spatial distribution of the electric field[15,22], the magnitude of nonlinear susceptibilities, and the influence of quantum effects[21], the angstrom-scale plasmonic gap structure plays decisive roles in enabling and enhancing these near-field nonlinear optical processes.

Finally, it should be noted that, as theoretically predicted by Luca and Ciracì[23], the field-effect modulation of surface charge density may also contribute to the field-induced modulation of our TE-SHG/TE-SFG for the bare the Au (111) substrate. They proposed that by applying an external electrostatic field to a slab of heavily doped semiconductors, the density of free electrons in a very small region at the top surface is modulated and the drastic enhancement of free electron nonlinear optical responses becomes possible. They further predicted that such an effect would become more pronounced in a nanopatterned structure that supports plasmonic excitation and two orders of magnitude boost of free electron nonlinear optical response could be achieved. Their prediction is consistent with our experimental results for the angstrom-scale plasmonic gap, in which nonlinear optical signals were drastically enhanced in response to the application of static electric field (≤ 1 V). Therefore, we currently consider that their proposed mechanism would contribute to the observed bias-dependent giant TE-SHG modulation for the bare Au(111) surface (Fig. S13). Although a comprehensive theoretical analysis is beyond the scope of this work, further investigation based on their theoretical framework and a quantitative comparison with our experimental results could substantiate this hypothesis, thereby unveiling a novel nonlinear electrophotonic modulation mechanism in quantum free-electron systems.

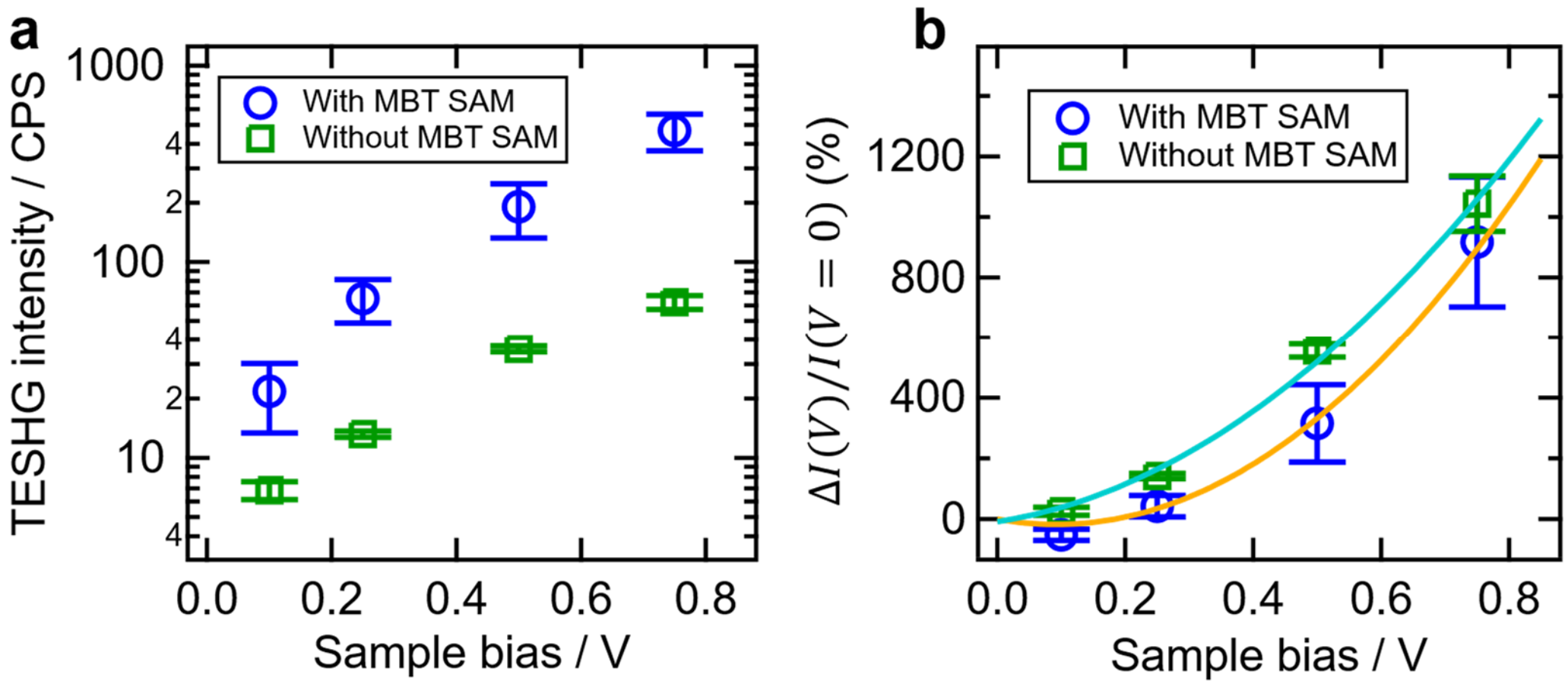

**Fig. S13 | TE-SHG intensity with and without SAM. a**, Sample bias dependence of absolute TE-SHG intensity obtained for Au(111) surface with (blue circles) and without (green squares) adsorption of MBT SAM. **b**, The sample bias-dependent change in the TE-SHG intensity ($\Delta I(V)$) normalized by the signal intensity at $V = 0$ V ($I(V = 0)$). $I(V = 0)$ was obtained by averaging the intensities at $V = -0.1$ V and $V = 0.1$ V. Blue circles and green squares represent the data for Au(111) surface with and without adsorption of MBT SAM, respectively. The orange and light blue curves are the results of the curve fitting with quadratic functions conducted for the data indicated by blue circles and green curves, respectively. The error bars in **a** and **b** represent the standard errors of each data point determined by repeating the same voltage-dependent experiments.

## 11. Sample bias dependence of TE-SHG under constant tunneling current.

In the bias-dependent TE-SHG measurements presented in Fig. 3 in the main text, we varied not only the applied bias but also the tunneling current setpoint in the range of 40–1750 pA, allowing us to prevent the variation in the tip-substrate distance and maintain the field enhancement strength (see Table S1 for the explicit setpoint values). To check whether this current tuning influenced the observed clear quadratic bias dependence and ~2000%/V modulation of TE-SHG intensity, we performed control bias-dependent TE-SHG measurements under a constant tunneling current setpoint (Fig. S14). As shown in Fig. S14b, even in this constant-current condition, the TE-SHG intensity exhibited quadratic bias dependence, and the large modulation depth (~2000 %/V) was still observed. These results indicate that the giant near-field nonlinear electrophotonic effects occur regardless of the tunneling current variations, demonstrating that the contributions of tunneling-current–driven processes, such as hot-carrier generation, inelastic scattering, impact ionization, photoluminescence, or electroluminescence, are negligibly small.

It should be noted that in this control experiment under the constant current mode, increasing the applied bias from 0.1 V to 1 V led to an elongation of the tip–substrate distance from ~4 Å to ~6 Å (Fig. S14a). Based on the previous classical electrodynamic simulations, it is expected that the near-field enhancement strength significantly decreases during this tip–substrate distance elongation[16,20,24–27]. Nevertheless, the TE-SHG intensity markedly increased with the applied bias, and the pronounced quadratic bias dependence with a modulation depth of ~2000%/V was still observed under the constant current mode (Fig. S14b). This result indicates that the expected distance-dependent variations in the field enhancement strength predicted by classical electromagnetic theory are essentially suppressed. This deviation from classical behavior can be attributed to quantum plasmonic quenching effects: at the gap distances of < 1 nm, the influences of quantum mechanical phenomena, such as electron spill-out from the metal surface and the overlap of electronic wavefunctions across the gap, begin to play roles and suppress the classically predicted field enhancement[15–21]. Particularly, in the gap distance range of 4–7 Å, these quantum suppression effects and the classically expected enhancement nearly cancel each other, making the overall field enhancement effectively independent of the tip–substrate distance[15–21]. Therefore, in the constant-current experiment shown in Fig. S14, not only current-induced effects but also distance-dependent variations in field enhancement strength can be reasonably excluded. This allows us to clearly attribute the observed TE-SHG modulation to the voltage-induced nonlinearity represented by the $\chi^{(3)}E_{\mathrm{DC}}$ term.

Notably, further reduction of the tip–substrate distance below ~4 Å leads to a regime where quantum plasmonic quenching becomes dominant, resulting in a steep decrease in the electric field enhancement factors[15–21]. In contrast, the 4–7 Å regime represents a critical crossover region just before this steep decline, where the electric field enhancement is maximized[15–21]. Due to this compensation, the overall field enhancement factor remains nearly constant at its maximum value throughout this gap-size range, rendering it effectively independent of tip–substrate distance from 4 Å to 7 Å. Thus, the 4–7 Å distance range not only offers optimal field enhancement conditions for both TE-SHG and TE-SFG processes, but also provides an ideal regime for exclusively probing the intrinsic bias dependence of the giant electrophotonic response, distinct from variations in the near-field enhancement.

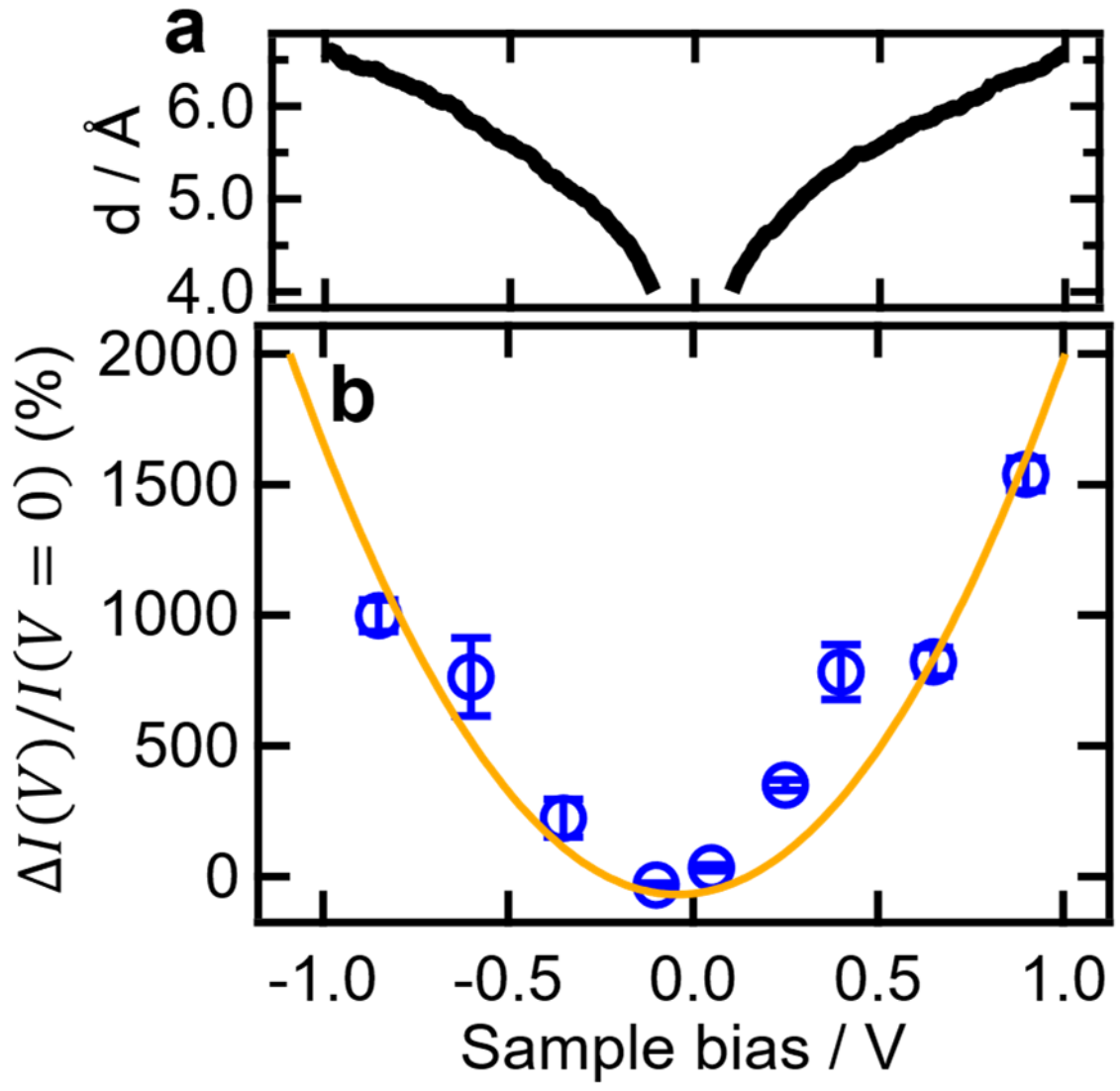

**Fig. S14 | Sample bias dependence of TE-SHG under constant tunneling current. a.** Sample bias dependence of the tip–substrate distance ($d$) with respect to the distance at feedback loop parameters of $V$ = 0.1 V and $I_t$ = 500 pA. The measurement was performed at a constant tunneling current of 500 pA. **b.** The voltage dependence of TE-SHG intensity under constant tunneling current (500 pA).

**12. Voltage dependent tip-enhanced nonlinear optical responses observed for different tips.**

We checked the reproducibility of the voltage dependence of TE-SHG and TE-SFG processes by measuring the intensities of TE-SHG and TE-SFG using three different tips shown in Fig. S15a–c. Although the slight differences in the apex structures could give rise to differences in the electrostatic field distribution and hence the fewfold differences in the degree of bias dependent signal enhancement, we successfully confirmed that these tips exhibited strong plasmonic enhancement of nonlinear optical responses and similar voltage-dependent enhancement behaviors emerged (Fig. S15d–f). This ensures not only the reproducibility of our results but also the generality of our concepts of giant electrical modulation of nonlinear optical processes based on an angstrom-scale plasmonic gap structure.

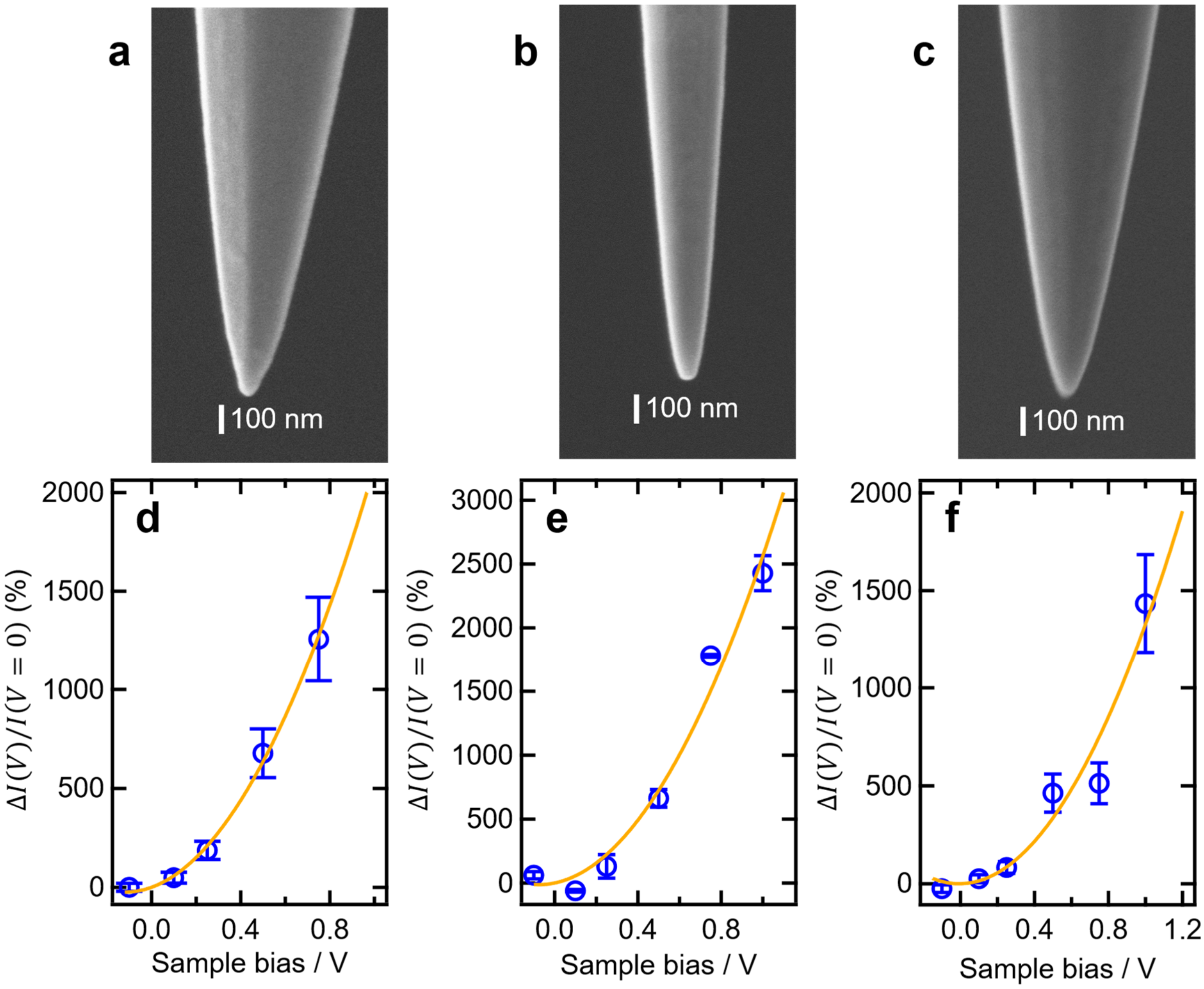

**Fig. S15 | Reproducibility of giant electrical enhancement of nonlinear optical processes with different tips. a–c.** Scanning electron micrographs of Au tips. These tips were independently fabricated and used in the experiments to ensure the reproducibility of giant electrical enhancement of nonlinear optical processes. **d–f.** The voltage-dependent changes in the TE-SHG (**d**) and TE-SFG (**e, f**) output ($\Delta I(V)$) normalized by the signal intensities at $V = 0$ V ($I(V=0)$). The orange curves are the result of curve fitting with a quadratic function. The data in (**d**), (**e**), and (**f**) were obtained by using the tips shown in (**a**), (**b**), and (**c**), respectively.

**13. Field enhancement mechanisms governing TE-SHG and TE-SFG processes—Broadband optical response spanning infrared to visible region—**

As discussed in the main text, we observed giant electrophotonic effects characterized by the quadratic voltage dependence and ~2000%/V modulation depth in TE-SHG and TE-SFG intensities. Since these nonlinear optical processes involve infrared-to-visible drastic frequency conversion between the incoming and outgoing light, a spectrally broad plasmonic enhancement that can simultaneously affect such separated frequencies is critical for realizing efficient generation of TE-SHG and TE-SFG. In the main text, however, we omitted the detailed description of the spectral properties of field enhancement strength in order to simplify the mathematical expression and exclusively focus on the effect induced by the electrostatic fields. Thus, in this section, we provide an overview of the electric-field enhancement mechanisms governing TE-SHG and TE-SFG processes by explicitly incorporating the plasmonic filed enhancement into the discussion[22,28].

SHG involves two-photon excitation (generation of nonlinear polarization $P^{(2)}$) and one-photon emission (radiation from $P^{(2)}$) processes. The excitation and radiation processes are simultaneously amplified at the angstrom-scale gap in STM through plasmonic field enhancement, leading to a substantial increase in SHG signals. Moreover, the third-order field-induced effect driven by the electrostatic field within the angstrom-scale gap ($E_{\mathrm{DC}}$) contributes to the modulation of the SHG signals. Consequently, the overall TE-SHG intensity ($I_{\mathrm{TESHG}}$) can be described by using the incident field enhancement factor ($K_{\mathrm{gap}} \equiv E_{\mathrm{gap}}/E_0$), the enhanced emission efficiency from $P^{(2)}$ ($L_{\mathrm{gap}}$), and second- and third-order nonlinear optical susceptibility ($\chi^{(2)}$ and $\chi^{(3)}$) as follows[22]:

$$I_{\mathrm{TESHG}}(2\omega) \propto \left|\chi^{(2)} + \chi^{(3)} E_{\mathrm{DC}}\right|^2 \left|K_{\mathrm{gap}}(\omega) E_0\right|^4 \left|L_{\mathrm{gap}}(2\omega)\right|^2, \tag{S2}$$

where $\omega$ represents the excitation frequency for TE-SHG. Equation (S2) provides a comprehensive expression for TE-SHG process that incorporates not only electrophotonic effect $(\chi^{(3)} E_{\mathrm{DC}})$ but also the spectral properties of the incident field enhancement ($K_{\mathrm{gap}}(\omega)$) and the signal emission efficiency ($L_{\mathrm{gap}}(2\omega)$). By denoting $\left|K_{\mathrm{gap}}(\omega) E_0\right|^2$ as $I_{\mathrm{gap}}$ and omitting $L_{\mathrm{gap}}(2\omega)$ term for simplicity, we arrive at equation (1) in the main text, representing a simplified expression where the electrophotonic effect on TE-SHG is more specifically featured. Here, to understand the fundamental mechanism governing TE-SHG process, we focus on the frequency-dependent terms in equation (S2) rather than the electrophotonic term. Since both incident excitation light and emitted SHG light in our experiments are non-resonant with electronic/vibrational transitions of MBT molecules and gold, we can assume that the frequency dependences of $\chi^{(2)}$ and $\chi^{(3)}$ are small. Therefore, the overall frequency profile of TE-SHG is approximated as

$$I_{\mathrm{TESHG}}(2\omega) \propto \left|K_{\mathrm{gap}}(\omega)\right|^4 \left|L_{\mathrm{gap}}(2\omega)\right|^2. \tag{S3}$$

Equation (S3) indicates that the spectral properties of TE-SHG can be generally described by considering the field enhancement factor ($K_{\mathrm{gap}}(\omega)$) and the enhanced emission efficiency ($L_{\mathrm{gap}}(2\omega)$). Therefore, the underlying enhancement mechanism of the TE-SHG process can be understood by examining $K_{gap}(\omega)$ and $L_{gap}(2\omega)$.

The procedure for electromagnetic field simulations was described elsewhere[22,28]. Briefly, the finite-difference time-domain (FDTD) method[29–31] was adopted with commercial software (Lumerical FDTD, Ansys). The system investigated in the simulation consists of a gold tip positioned above a gold substrate in vacuum, representing the nanogap in our STM. The refractive index of gold was taken from the experimental values of Olmon *et al.*[32] Perfectly matched layer boundary conditions were used in all simulations to absorb all outgoing waves and eliminate light reflection. To evaluate the spectral properties of incident field enhancement ($K_{gap}$), we placed a monitor at the midpoint between the tip apex and the substrate surface to measure the electromagnetic field strength. A $p$-polarized Gaussian beam source with a waist of 2 μm was used to illuminate the nanogap at an incident of 55°.

To evaluate the radiation efficiency ($L_{gap}$), an oscillating dipole source perpendicular to the gold substrate, representing the nonlinear polarizations generated within the gap, was placed at the same position as the monitor for $K_{gap}$. The selection of the dipole source is verified based on our recent demonstration[10]: we experimentally and theoretically revealed that when the tip apex size is on the order of ~50 nm as in the present study (Fig. 1b in the main text), the generated nonlinear polarizations are highly dipole in nature, with no significant contributions from quadrupoles or higher-order multipoles[10]. Radiated electromagnetic field from the dipole was monitored at a position where the lateral and vertical distances from the dipole were 3000 and 2100 nm, respectively, corresponding to the reflection angle of 55° employed in our experiments. Note that nonlocal or quantum-corrected models[15–21] were not explicitly considered in our simulations. Moreover, since the near-field enhancement factor within the tip–substrate gap was revealed to be essentially unaffected by the STM bias of ~1 V applied across a tip–substrate junction maintained at constant distance[33], the influence of the applied bias was not incorporated in our FDTD simulation.

In the following, we review previously reported spectral characteristics of the field enhancement factor and radiation efficiency[22,28] to facilitate a clear understanding of the broadband IR-to-visible near-field enhancement that is essential for the TE-SHG and TE-SFG processes. Initially, we started by modeling an STM tip as a nanosphere without a shaft[22,27,34–38] (Fig. S16a), and the wavelength dependences of $K_{gap}$ and $L_{gap}$ calculated for this nanosphere-substrate system is shown in Fig. S16b and c, respectively. The spectra have a single enhancement band in the visible region. This result aligns with a typical signature of the gap-mode plasmon excitation localized between the nanogap.[35,39–41] However, in the infrared region, both of the $K_{gap}$ and $L_{gap}$ values seem to be too small to explain the enhancement of infrared-light-excited TE-SHG and TE-SFG. This implies that micrometer-scale tip shafts, which were not considered in the nanosphere-substrate system, play an important role in the enhancement of the infrared region.

To clarify the influence of the micrometer-scale tip shafts, we then performed more realistic FDTD simulations, in which an STM gap structure was modeled by a rounded cone-shaped gold tip with a longitudinal shaft (tip length: $l$) placed above a flat gold substrate (Fig. S16d)[22]. The opening angle and curvature radius of the tip apex were 30° and 50 nm, respectively. Figure S16e shows the $K_{\text{gap}}$ spectra calculated for various tip lengths. For short tips ($l \leq 150$ nm), the enhancement remained localized in the visible region, driven by the gap-mode plasmon resonance described above. In contrast, as the tip length is extended (particularly $l \geq 600$ nm), the broadband enhancement profile covering the broad near- and mid-IR region emerged. We confirmed that the calculation converged at $l = 15000$ nm and the broadband electric field enhancement remains at this shaft length. The results obtained for the long tips correspond to the actual experimental situation, where the Au tips with micrometer-scale lengths (Fig. 1b in the main text and Fig. S15a–c) were used. Therefore, we can conclude that the field enhancement at the nanogap is highly effective over broad wavelength range encompassing visible, near-IR and mid-IR region. This drastic enhancement in the infrared region is a clear manifestation of the influence of the long tip shafts, whose origin can be attributed to the antenna effect caused by the collective oscillation of electrons over the entire tip[42–45].

On the other hand, in the radiation process, the effect of the tip shafts is not pronounced. As shown in Fig. S16f, the spectra of $L_{\text{gap}}$ were still limited to a single band in the visible domain regardless of the tip length $l$. Although the variations in the tip length cause slight differences in the strength and shape of the $L_{\text{gap}}$ spectra, the wavelength range of efficient radiation from nanocavities is predominantly determined by gap-mode plasmons. This is because the time-averaged power of the vacuum propagating electromagnetic field radiated by the oscillating polarization is proportional to $\lambda^{-4}$ and steeply decreases with wavelength, making the contributions from the antenna effects in the IR region less dominant. Note that the center wavelengths of TE-SHG and TE-SFG outputs in this study (750 nm and 785 nm, respectively) are covered by the gap-mode resonance band in the visible region, though they are located in the longer-wavelength edge. Prior studies demonstrated that gap-mode plasmons in this longer wavelength range can be well approximated as dipolar modes[27]. Therefore, the radiation process is mediated predominantly by dipolar gap-mode plasmons.

These spectral characteristics of $K_{\text{gap}}$ and $L_{\text{gap}}$ in the long tips are key to understand the mechanism of infrared-excited TE-SHG process. Figure S17a shows the wavelength dependence of the TE-SHG intensity calculated using equation (S3) and $K_{\text{gap}}$ and $L_{\text{gap}}$ spectra obtained for $l = 15000$ nm. This spectral profile indicates that the TE-SHG process is highly efficient over broad near-IR range encompassing $\lambda \geq 1100$ nm region. This spectrally broad effectiveness of TE-SHG is the consequence of the fact that the broadband enhancement of incident light in the near-to-mid-infrared region ($K_{\text{gap}}$) and efficient radiation in the visible-to-near-infrared region ($L_{\text{gap}}$) effectively cover the excitation and SHG radiation wavelength ranges, respectively. Thus, the SHG enhancement arises from the simultaneous amplification at $\omega$ (excitation) and $2\omega$ (radiation) caused by the concerted operation of two distinct enhancement mechanisms: the antenna effects caused by micrometer-scale tip shafts enhance infrared excitation, while localized gap-mode plasmons intensify the radiation of second harmonics.

It should also be remarked that a similar discussion can be applied to the TE-SFG process. In analogy with equation (S3), the frequency profile of TE-SFG can be described by the product of the enhancement factor and radiation efficiency:

$$I_{\mathrm{TESFG}}(\omega_{\mathrm{SFG}}) \propto \left|K_{\mathrm{gap}}(\omega_1)\right|^2 \left|K_{\mathrm{gap}}(\omega_2)\right|^2 \left|L_{\mathrm{gap}}(\omega_{\mathrm{SFG}})\right|^2, \tag{S4}$$

where $I_{\mathrm{TESFG}}$ is the output TE-SFG intensity; $\omega_1$ and $\omega_2$ represent the two different frequencies of excitation light for TE-SFG; and $\omega_{\mathrm{SFG}}$ represents the sum of $\omega_1$ and $\omega_2$. Similarly to the case of TE-SHG, equation (S4) allows us to predict the wavelength dependence of TE-SFG intensity (Fig. S17b), exhibiting the broadband effectiveness spanning not only near-IR but also mid-IR region. Therefore, the concerted effect of the antenna effects in the infrared region and the gap-mode plasmon in the visible region also governs the enhancement mechanism of TE-SFG process.

To further improve the TE-SHG and TE-SFG efficiencies, it is essential to tailor the enhancement spectrum such that the value of $\left|K_{\mathrm{gap}}K_{\mathrm{gap}}L_{\mathrm{gap}}\right|^2$ terms in equations (S3) and (S4) are maximized. As discussed above and demonstrated in more detail in our previous work[22], the spectral profiles of $K_{\mathrm{gap}}$ and $L_{\mathrm{gap}}$ are highly sensitive to both nanometer- and micrometer-scale tip geometry. Specifically, the nanometer-scale curvature of the tip apex primarily determines the gap-mode plasmon resonance in the visible region, whereas the micrometer-scale surface geometry of the tip shaft governs the broader field enhancement at the IR region through the antenna effect[22]. Therefore, to improve the field enhancement strengths at both the visible and IR regions, simultaneous control of these different-scale structures through nanoscale adjustment of the apex curvature[46] and the introduction of grating patterns on the micrometer-scale tip shaft[47,48] is essential, leading to further improvement of the TE-SHG/TE-SFG efficiencies. Such fine control of tip geometries would be achieved by exploiting more sophisticated tip processing techniques, such as field-directed sputter sharpening for the tip apex[49] and focused ion beam processing for the tip shaft[50]. We believe that implementing these strategies to improve the overall near-field nonlinear optical efficiencies represents an important research direction for our future work, potentially leading to novel strategies to improve the conversion efficiency of near-field nonlinear optical processes.

Finally, to show the influence of the presence of the SAM layer on the field enhancement spectra, we performed additional electromagnetic field simulations incorporating 6-Å-thick SAM layer within the tip–substrate gap with 1-nm distance (Fig. S18a). The refractive index of the SAM layer was set to be 1.2, which represents the typical value that has been used to calculate the optical responses of interfacial SAM layers[51–53]. As shown in Figs. S18b and c, the presence of the SAM layer does not significantly alter the spectral profiles of both field enhancement factor (Fig. S18b) and radiation efficiency (Fig. S18c). Therefore, although the above discussion on the field enhancement mechanisms is based on the vacuum gap, the same conclusion can be deduced even in the presence of the SAM layer within the gap.

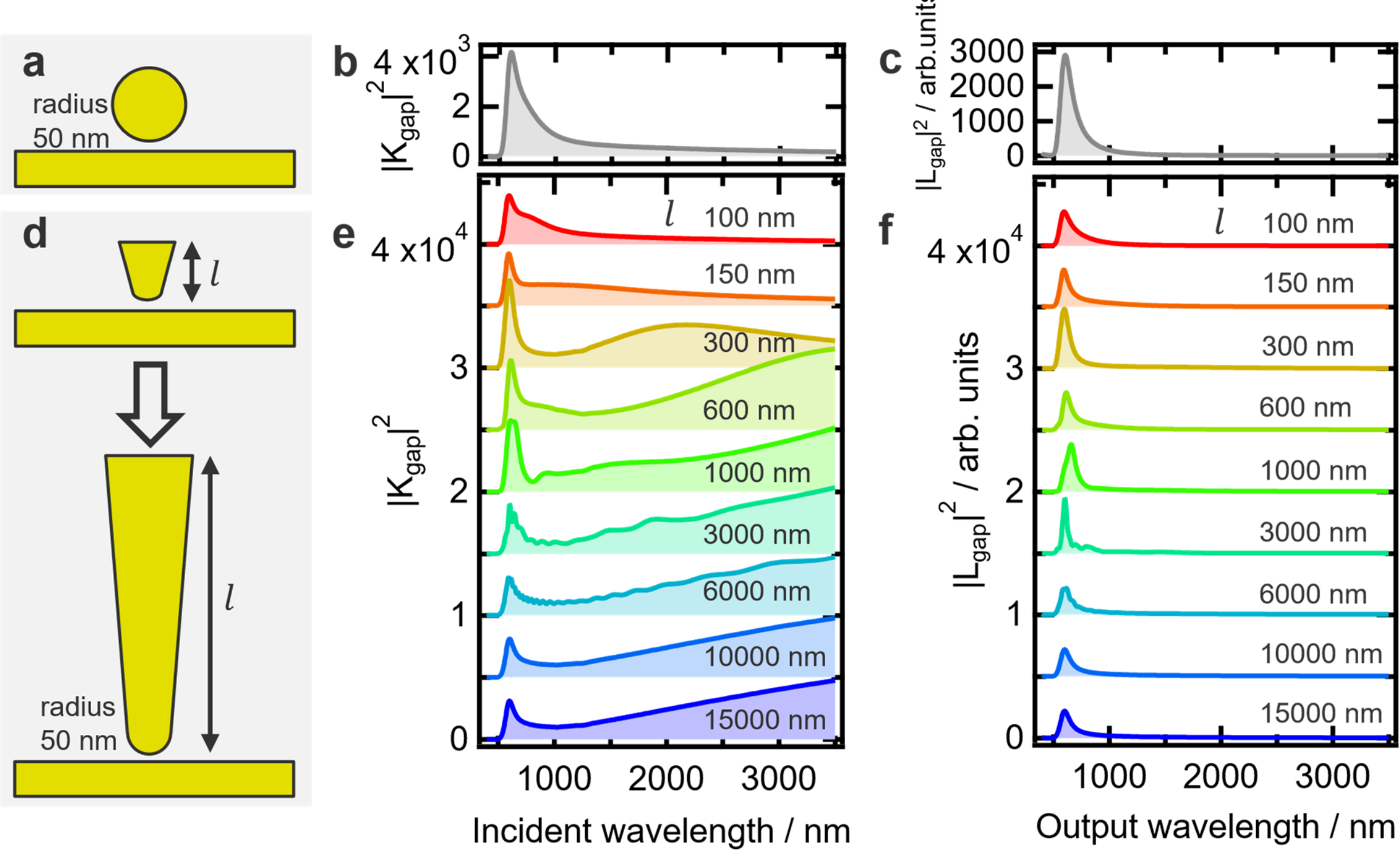

**Fig. S16 | Theoretical calculation of the field enhancement factor and the radiation efficiency revealing the mechanism of infrared-to-visible broadband nonlinear optical responses. a.** Schematic representation of nanosphere-substrate configuration. The radius of the nanosphere is 50 nm. **b.** $|K_{\text{gap}}|^2$ and **c.** $|L_{\text{gap}}|^2$ spectra of the nanogap in a nanosphere-substrate system calculated through the FDTD method. **d.** Schematic representation of tip-substrate configuration. A rounded cone tip with a 30° opening angle and 50 nm radius of curverture was adopted in the calculation. The tip length $l$ was changed from 100 nm to 15000 nm. **e, f**. Tip-length dependent (**e**) $|K_{\text{gap}}|^2$ and (**f**) $|L_{\text{gap}}|^2$ spectra of a tip–substrate nanocavity calculated through the FDTD method. The tip lengths are indicated in the figures and the tip–substrate distance $d$ was taken as 1 nm for all these calculations. The calculation in **e** and **f** converged at $l = 15000$ nm. The results obtained for longer tips corresponds to the actual experimental conditions where the micrometer-scale Au tips (Fig. S15a–c) were used.

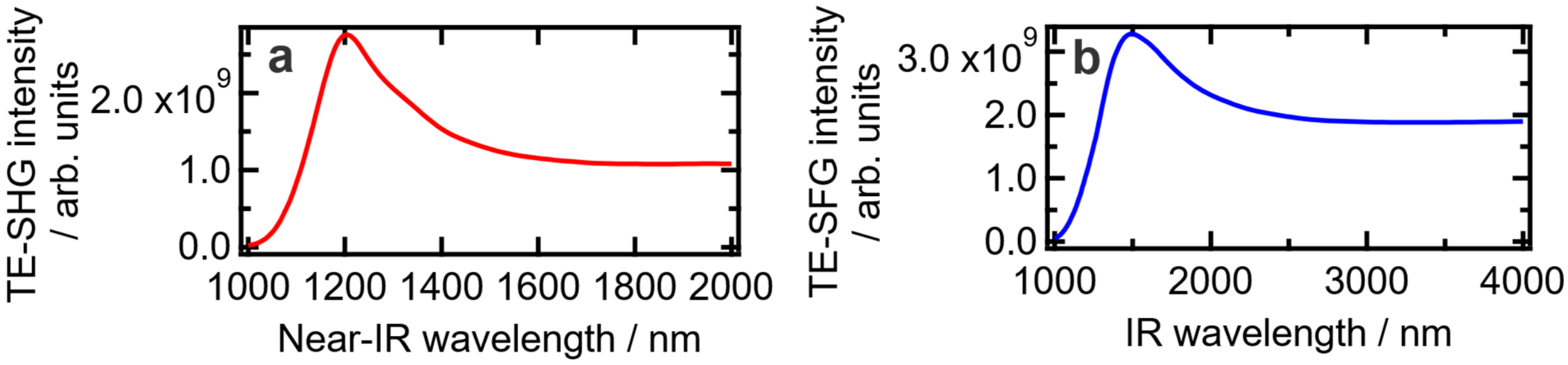

**Fig. S17 | Spectral properties of tip-enhanced second-order nonlinear optical processes.** The excitation wavelength dependence of TE-SHG (**a**) and TE-SFG (**b**) calculated based on equations (S3) and (S4), respectively. In calculating these spectral properties, $|K_{\mathrm{gap}}|^2$ and $|L_{\mathrm{gap}}|^2$ spectra obtained for $l = 15000\ \mathrm{nm}$ tip (Fig. S16e and f) were adopted. Note that the horizontal axis in b represents one of the two excitaiton wavelengths of SFG. Another excitation wavelength was fixed at 1033 nm, which corresponds to the central wavelength of near-IR excitation pulses used in our experiments.

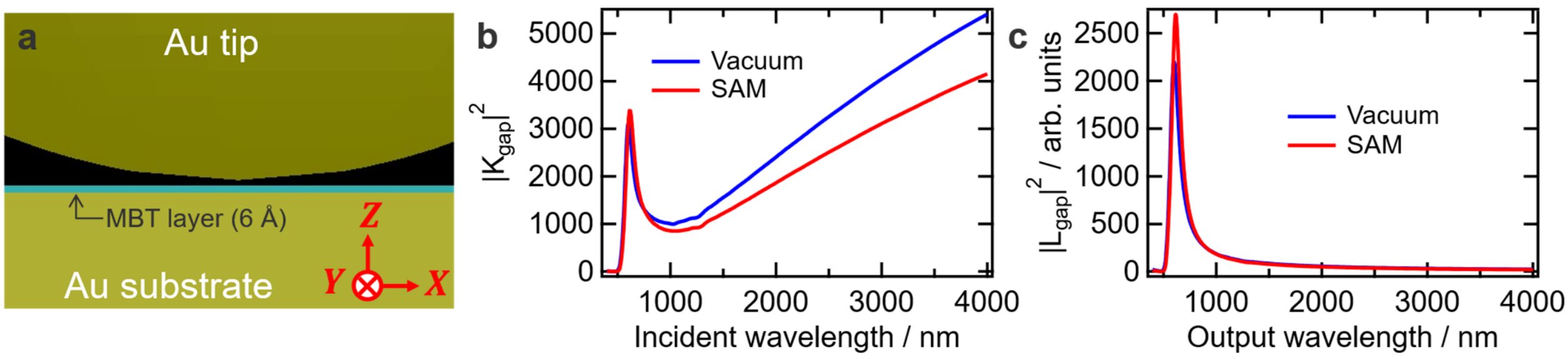

**Fig. S18 | Theoretical calculation of the near-field properties in the presence of the MBT layer within the gap. a.** Schematic representation of the gap region including the MBT layer with a thickness of 6 Å and a refractive index of 1.2 (the light blue region). The metal–to–metal separation between the tip apex and the substrate surface was 1 nm. **b, c**. The red curves represent the spectra of (**b**) the field enhancement factor ($|K_{\mathrm{gap}}|^2$) and (**c**) the radiation efficiency ($|L_{\mathrm{gap}}|^2$) calculated under the MBT-introduced gap configuration shown in panel (**a**). A rounded cone tip with a 30° opening angle, 50 nm radius of curverture, and 15000-nm length was adopted in the calculation. The calculated results obtained without incorporating the MBT layer (the blue curves in **b** and **c**) are also shown as a reference. Note that these blue curves are identical to the data at $l = 15000\ \mathrm{nm}$ shown in Fig. S16.

## 14. Spatial distribution of electric fields and charge density

In this section, we quantitatively estimate the spatial distribution of the electric field within the nanogap. Figure S19 displays the surface-normal and surface-parallel components of the electric field within the tip–substrate gap ($E_{\text{gap}}$), which were calculated through the FDTD simulation for a tip with a curvature radius of 50 nm. This tip radius is comparable to the size of the tip apex used in our experiments (Fig. 1b in the main text). Note that the tip–substrate gap was assumed to be vacuum, and the presence of the SAM layer was not incorporated. As shown in Fig. S19, the enhanced electric field is spatially non-uniform and distributed over the nanoscale region with a radius of ~20 nm. Furthermore, our theoretical analysis of the electrostatic field distribution within the gap using the image charge method[54] revealed that the electrostatic field across the gap ($E_{\text{DC}}$) also exhibits a similar spatial distribution (Fig. S20). Importantly, our recent experiments revealed that near-field second-order nonlinear optical signals originate from a nanoscale region with an area of a few tens of nanometers directly beneath the tip apex[10]. Since this result is consistent with theoretically predicted ~20-nm-scale lateral field enhancement areas (Figs. S19 and S20), the effective field strength averaged over this ~20 nm region mainly contributes to the generation of near-field second-order nonlinear optical signals.

Based on the spatial distributions and relative intensities of the surface-normal and surface-parallel components, we can directly identify the main nonlinear susceptibility tensor components contributing to the near-field nonlinear generation and the resultant polarization state of emitted light. As shown in Fig. S19, the surface-normal (*Z*-directed) near-field component ($E_Z$) is more than one order of magnitude stronger than the surface-parallel field component, indicating that only $E_Z$ dominantly contributes to nonlinear optical processes within the gap. In this case, among the 27 tensor components of the second-order nonlinear susceptibility, only $\chi^{(2)}_{XZZ}$, $\chi^{(2)}_{YZZ}$, or $\chi^{(2)}_{ZZZ}$ can in principle contribute to the second-order nonlinear polarization within the tip–substrate gap. When the cylindrical symmetry about the central axis of the tip is satisfied, the contributions of $\chi^{(2)}_{XZZ}$ and $\chi^{(2)}_{YZZ}$ are eliminated, leaving $\chi^{(2)}_{ZZZ}$ as the dominant component[55]. Therefore, the generated second-order nonlinear polarization is mainly oriented along the *Z*-axis, and the corresponding emitted field is *p*-polarized.

We would like to note that, consistent with this argument, the dominance of $\chi^{(2)}_{ZZZ}$ was also confirmed experimentally in our recent work[10]. By measuring the vibrationally resonant TE-SFG for methyl vibrations in MBT molecules, we revealed that the methyl antisymmetric stretching modes exhibited negative $\text{Im}\big(\chi^{(2)}\big)$ spectra. Based on the considerations of the explicit values of hyperpolarizability tensors and angular distributions of MBT molecules, $\chi^{(2)}_{ZZZ}$ associated with the methyl antisymmetric modes exhibits negative imaginary part, whereas other hyperpolarizability tensor components, such as $\chi^{(2)}_{XXZ}$ and $\chi^{(2)}_{YYZ}$, are characterized by positive imaginary parts. Therefore, the observation of the negative $\text{Im}\big(\chi^{(2)}\big)$ indicates that $\chi^{(2)}_{ZZZ}$ component predominantly contributes to the generation of the tip-enhanced second-order nonlinear polarization. Therefore, not only theoretical analysis but also our experimental results[10] consistently corroborates the predominance of $\chi^{(2)}_{ZZZ}$, indicating that near-field nonlinear signal radiation is dominantly *p*-polarized.

Similar to the second-order case, the third-order polarization is also oriented along the *Z*-axis. In addition to the oscillating near-field induced by the incident excitation light (Fig. S19), the electrostatic field generated through the voltage application across the tip–substrate gap is also dominated by its *Z*-component (Fig. S20). Therefore, by analogy with the second-order susceptibility, only $\chi^{(3)}_{ZZZZ}$ component is active among all third-order susceptibility tensor components. This ensures that the third-order nonlinear polarization in EFISH is also aligned along the *Z*-axis and that the corresponding emitted field is *p*-polarized.

Finally, to elucidate the physical origin of the near-field enhancement at the tip–substrate gap, we calculated the charge density distribution in the gap region by applying Gauss's law ($\nabla \cdot E \propto \rho$) to the spatial electric field distribution shown in Fig. S19. Figure S21a displays the charge density distribution on the *XZ* plane ($Y = 0$ nm) under the 3280-nm excitation, showing that charge neutrality is preserved within the bulk due to screening effects and charges are localized at the metal surfaces. Similar charge distribution is predicted for the near-IR (1033 nm) excitation conditions (Fig. S21b). Notably, under these mid- and near-IR excitation conditions, charges of opposite signs are simply distributed across the tip and substrate without forming any multipole-like complex charge distribution. Therefore, the plasmon mode generated under the near-IR irradiation can be regarded as strongly dipolar in nature.

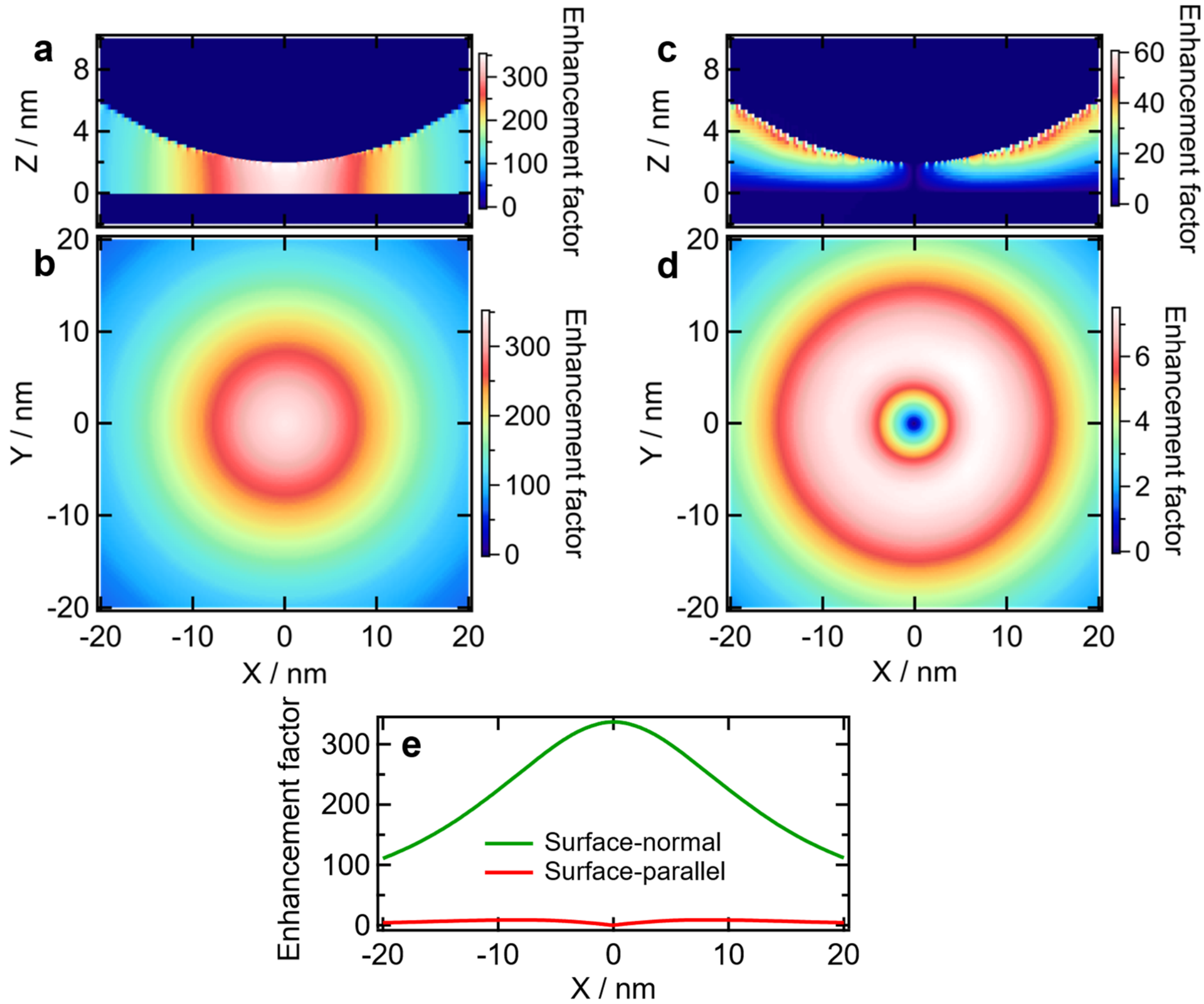

**Fig. S19 | Spatial distributions of the field enhancement factor $K_{gap}$.** The theoretical predictions of (**a, b**) surface-normal and (**c, d**) surface-parallel electric field components calculated through the

FDTD method are displayed. The model employs a rounded gold cone with a 30° opening angle, 50-nm apex radius, and 15000-nm length. The tip–substrate distance was set to be 2 nm. Panels (**a**) and (**c**) show the field in the *XZ* plane ($Y$ = 0 nm), while panels (**b**) and (**d**) show the field in the *XY* plane ($Z$ = 0.5 nm). The wavelength of incident light was 3280 nm. The coordinates ($X$, $Y$) = (0 nm, 0 nm) represent the position of minimum tip–substrate distance. (**e**) One dimensional spatial profiles of $K_{\mathrm{gap}}$ along the $X$-axis ($Y$ = 0 nm and $Z$ = 0.6 nm). Green and red curves represent the surface-normal and surface-parallel field components, respectively. The green curve is identical to that in Figure 5c in the main text.

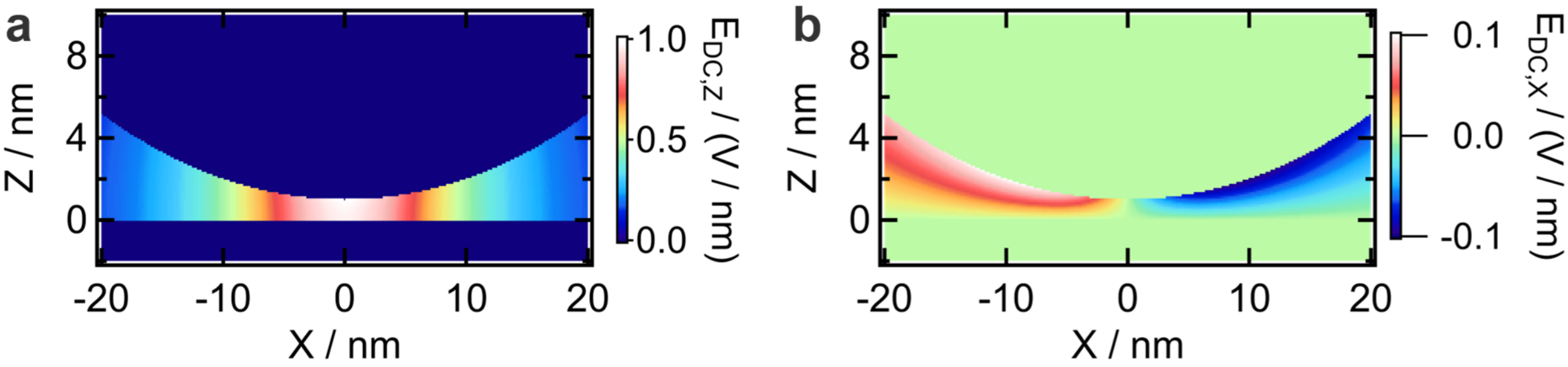

**Fig. S20 | Simulated spatial distributions of the electrostatic field $E_{\mathrm{DC}}$ under 1 V of voltage application across the tip-substrate nanogap.** The calculation was performed through the image charge method[54]. The surface-normal (a) and surface-parallel (b) electrostatic field components are plotted in the XZ plane (Y = 0 nm). In the calculation, the tip was approximated by a nanosphere with a 50-nm apex radius. The tip–substrate distance was set to be 1 nm. The coordinates (X, Y) = (0 nm, 0 nm) represent the position of minimum tip–substrate distance.

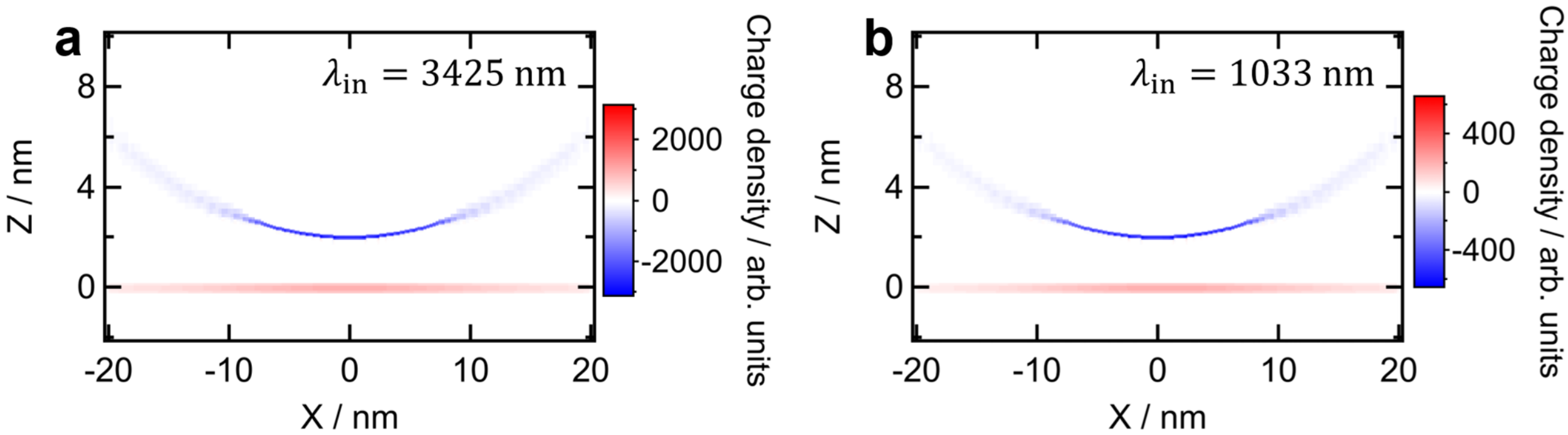

**Fig. S21 | Charge distribution across the tip and substrate on the *XZ* plane ($Y$ = 0 nm) under the IR excitation.** These charge densities were obtained by applying Gauss's law ($\nabla \cdot E \propto \rho$) to the spatial electric field distribution. Panels (**a**) and (**b**) represent the Fourier component corresponding to 3280-nm and 1033-nm excitation, respectively. The bulk region is electrically neutral due to the screening effects, and charges with opposite signs are accumulated at the metal surfaces.

## 15. Comparison between tip plasmon and junction plasmon

In the main text and Supplementary Section 5, we discussed the SHG and SFG emission behaviors when the tip and substrate are retracted by 30 nm. Importantly, although junction plasmons are not excited in these retracted conditions, the tip apex is still irradiated by the excitation light, potentially exciting plasmons localized at the tip apex. To evaluate the effects of these tip plasmons, we specifically investigated the field enhancement behavior in the absence of angstrom-scale gap structures by performing additional FDTD calculation.

In the additional calculation, we expanded the tip–surface distance to 30 nm, which corresponds to the retracted condition in our experiments and calculated the spectra of the field enhancement factor $\left(|K_{\mathrm{gap}}|^2\right)$ and the emission efficiency $\left(|L_{\mathrm{gap}}|^2\right)$ (Fig. S22). For calculating the $|K_{\mathrm{gap}}|^2$ derived from the tip plasmon, the monitor for the electromagnetic field was placed 0.5 nm below the tip apex. As a result, the electric field enhancement (Fig. S22a) and emission efficiency (Fig. S22b) under the 30-nm-retracted condition were found to be more than one order of magnitude smaller than those in an angstrom-scale tip–substrate plasmonic gap. These values should be too weak to produce a detectable enhancement of nonlinear optical signals from nanoscale tip apex, thereby allowing us to safely disregard the contribution from tip plasmon-enhanced signals in our experiments.

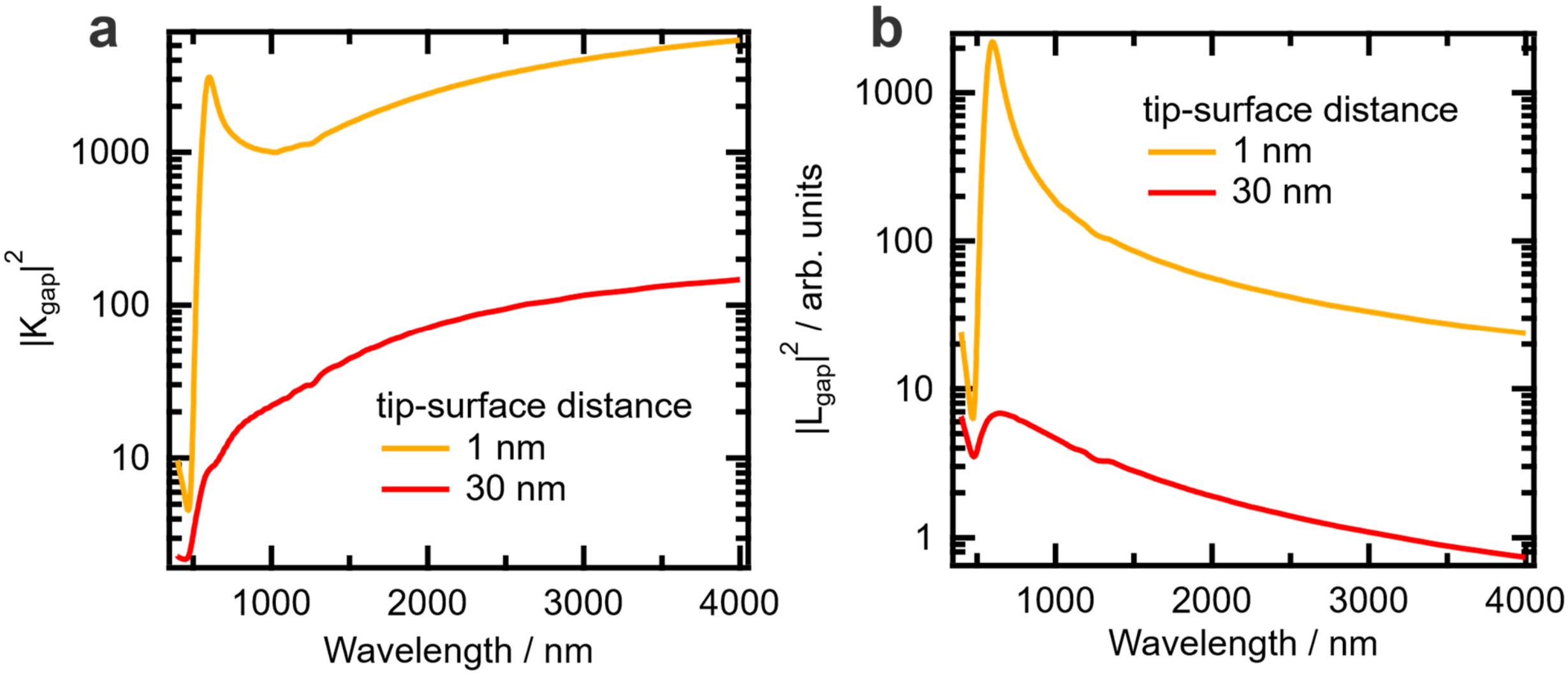

**Fig. S22 | Calculation of field enhancement caused by a tip plasmon.** The spectra of field enhancement factor $|K_{\mathrm{gap}}|^2$ (**a**) and emission efficiency $|L_{\mathrm{gap}}|^2$ (**b**) calculated for 1-nm (orange) and 30-nm (red) tip-substrate gap distances. The orange curves in **a** and **b** are identical to those shown in Fig. S16e and f obtained when $l = 15000$ nm, respectively.

## References

1. Hong, M., Yokota, Y., Hayazawa, N., Kazuma, E. & Kim, Y. Homogeneous Dispersion of Aromatic Thiolates in the Binary Self-Assembled Monolayer on Au(111) via Displacement Revealed by Tip-Enhanced Raman Spectroscopy. *J. Phys. Chem. C* **124**, 13141–13149 (2020).
2. Yang, G. & Liu, G. New Insights for Self-Assembled Monolayers of Organothiols on Au(111) Revealed by Scanning Tunneling Microscopy. *J. Phys. Chem. B* **107**, 8746–8759 (2003).
3. Yokota, Y. *et al.* Systematic Assessment of Benzenethiol Self-Assembled Monolayers on Au(111) as a Standard Sample for Electrochemical Tip-Enhanced Raman Spectroscopy. *J. Phys. Chem. C* **123**, 2953–2963 (2019).
4. Okabayashi, N., Konda, Y. & Komeda, T. Inelastic Electron Tunneling Spectroscopy of an Alkanethiol Self-Assembled Monolayer Using Scanning Tunneling Microscopy. *Phys. Rev. Lett.* **100**, 217801 (2008).
5. Bumm, L. A., Arnold, J. J., Dunbar, T. D., Allara, D. L. & Weiss, P. S. Electron Transfer through Organic Molecules. *J. Phys. Chem. B* **103**, 8122–8127 (1999).
6. Fujii, S., Ziatdinov, M., Higashibayashi, S., Sakurai, H. & Kiguchi, M. Bowl Inversion and Electronic Switching of Buckybowls on Gold. *J. Am. Chem. Soc.* **138**, 12142–12149 (2016).
7. Yang, B. *et al.* Chemical Enhancement and Quenching in Single-Molecule Tip-Enhanced Raman Spectroscopy. *Angew. Chem. Int. Ed.* **62**, e202218799 (2023).
8. Chen, C. J. *Introduction to Scanning Tunneling Microscopy*. (Oxford University Press, 1994).
9. Seo, K. & Borguet, E. Potential-Induced Structural Change in a Self-Assembled Monolayer of 4-Methylbenzenethiol on Au(111). *J. Phys. Chem. C* **111**, 6335–6342 (2007).
10. Takahashi, S. *et al.* Diffraction-Unlimited Tip-Enhanced Sum-Frequency Vibrational Nanoscopy. *arXiv:2509.09179 [physics.optics]*.
11. Domke, K. F., Zhang, D. & Pettinger, B. Toward Raman Fingerprints of Single Dye Molecules at Atomically Smooth Au(111). *J. Am. Chem. Soc.* **128**, 14721–14727 (2006).
12. Haiss, W., Lackey, D., Sass, J. K. & Besocke, K. H. Atomic resolution scanning tunneling microscopy images of Au(111) surfaces in air and polar organic solvents. *J. Chem. Phys.* **95**, 2193–2196 (1991).
13. Barth, J. V., Brune, H., Ertl, G. & Behm, R. J. Scanning tunneling microscopy observations on the reconstructed Au(111) surface: Atomic structure, long-range superstructure, rotational domains, and surface defects. *Phys. Rev. B* **42**, 9307–9318 (1990).
14. Fu, Y., Zhang, P., Verboncoeur, J. P. & Wang, X. Electrical breakdown from macro to micro/nano scales: a tutorial and a review of the state of the art. *Plasma Res. Express* **2**, 013001 (2020).
15. Liu, S. *et al.* Inelastic Light Scattering in the Vicinity of a Single-Atom Quantum Point Contact in a Plasmonic Picocavity. *ACS Nano* **17**, 10172–10180 (2023).
16. Zuloaga, J., Prodan, E. & Nordlander, P. Quantum Description of the Plasmon Resonances of a Nanoparticle Dimer. *Nano Lett.* **9**, 887–891 (2009).
17. Savage, K. J. *et al.* Revealing the quantum regime in tunnelling plasmonics. *Nature* **491**, 574–577 (2012).

18. Esteban, R., Borisov, A. G., Nordlander, P. & Aizpurua, J. Bridging quantum and classical plasmonics with a quantum-corrected model. *Nat. Commun.* **3**, 825 (2012).
19. Zhu, W. & Crozier, K. B. Quantum mechanical limit to plasmonic enhancement as observed by surface-enhanced Raman scattering. *Nat. Commun.* **5**, 5228 (2014).
20. Marinica, D. C., Kazansky, A. K., Nordlander, P., Aizpurua, J. & Borisov, A. G. Quantum Plasmonics: Nonlinear Effects in the Field Enhancement of a Plasmonic Nanoparticle Dimer. *Nano Lett.* **12**, 1333–1339 (2012).
21. Zhu, W. *et al.* Quantum mechanical effects in plasmonic structures with subnanometre gaps. *Nat. Commun.* **7**, 11495 (2016).
22. Takahashi, S., Sakurai, A., Mochizuki, T. & Sugimoto, T. Broadband Tip-Enhanced Nonlinear Optical Response in a Plasmonic Nanocavity. *J. Phys. Chem. Lett.* **14**, 6919–6926 (2023).
23. De Luca, F. & Ciracì, C. Impact of Surface Charge Depletion on the Free Electron Nonlinear Response of Heavily Doped Semiconductors. *Phys. Rev. Lett.* **129**, 123902 (2022).
24. Downes, A., Salter, D. & Elfick, A. Finite Element Simulations of Tip-Enhanced Raman and Fluorescence Spectroscopy. *J. Phys. Chem. B* **110**, 6692–6698 (2006).
25. Chen, X. & Wang, X. Near-field thermal transport in a nanotip under laser irradiation. *Nanotechnology* **22**, 075204 (2011).
26. Esteban, R. *et al.* A classical treatment of optical tunneling in plasmonic gaps: extending the quantum corrected model to practical situations. *Faraday Discuss.* **178**, 151–183 (2015).
27. Martín-Jiménez, A. *et al.* Unveiling the radiative local density of optical states of a plasmonic nanocavity by STM. *Nat. Commun.* **11**, 1021 (2020).
28. Sakurai, A., Takahashi, S., Mochizuki, T. & Sugimoto, T. Tip-Enhanced Sum Frequency Generation for Molecular Vibrational Nanospectroscopy. *Nano Lett.* **25**, 6390–6398 (2025).
29. Kane Yee. Numerical solution of initial boundary value problems involving maxwell's equations in isotropic media. *IEEE Trans. Antennas Propagat.* **14**, 302–307 (1966).
30. Anees, A. & Angermann, L. Time Domain Finite Element Method for Maxwell's Equations. *IEEE Access* **7**, 63852–63867 (2019).
31. Teixeira, F. L. *et al.* Finite-difference time-domain methods. *Nat. Rev. Methods Primers* **3**, 75 (2023).
32. Olmon, R. L. *et al.* Optical dielectric function of gold. *Phys. Rev. B* **86**, 235147 (2012).
33. Jaculbia, R. B. *et al.* Single-molecule resonance Raman effect in a plasmonic nanocavity. *Nat. Nanotechnol.* **15**, 105–110 (2020).
34. Heilman, A. L., Hermann, R. J. & Gordon, M. J. Direct detection of gap mode plasmon resonances using attenuated total reflection-based tip-enhanced near-field optical microscopy. *J. Opt.* **22**, 095001 (2020).
35. Porto, J. A., Johansson, P., Apell, S. P. & López-Ríos, T. Resonance shift effects in apertureless scanning near-field optical microscopy. *Phys. Rev. B* **67**, 085409 (2003).
36. Krug, J. T., Sánchez, E. J. & Xie, X. S. Design of near-field optical probes with optimal field enhancement by finite difference time domain electromagnetic simulation. *J. Chem. Phys.* **116**,

10895–10901 (2002).

37. Madrazo, A., Nieto-Vesperinas, M. & García, N. Exact calculation of Maxwell equations for a tip-metallic interface configuration: Application to atomic resolution by photon emission. *Phys. Rev. B* **53**, 3654–3657 (1996).

38. Esteban, R., Vogelgesang, R. & Kern, K. Tip-substrate interaction in optical near-field microscopy. *Phys. Rev. B* **75**, 195410 (2007).

39. Li, G. *et al.* Plasmonic enhancement and directional emission for side-illumination tip-enhanced spectroscopy. *Opt. Commun.* **442**, 50–55 (2019).

40. Wei, Y., Pei, H., Sun, D., Duan, S. & Tian, G. Numerical investigations on the electromagnetic enhancement effect to tip-enhanced Raman scattering and fluorescence processes. *J. Phys.: Condens. Matter* **31**, 235301 (2019).

41. Futamata, M., Ishikura, M., Iida, C. & Handa, S. The critical importance of gap modes in surface enhanced Raman scattering. *Faraday Discuss.* **178**, 203–220 (2015).

42. Huth, F. *et al.* Resonant Antenna Probes for Tip-Enhanced Infrared Near-Field Microscopy. *Nano Lett.* **13**, 1065–1072 (2013).

43. Mastel, S. *et al.* Terahertz Nanofocusing with Cantilevered Terahertz-Resonant Antenna Tips. *Nano Lett.* **17**, 6526–6533 (2017).

44. Hermann, R. J. & Gordon, M. J. Quantitative comparison of plasmon resonances and field enhancements of near-field optical antennae using FDTD simulations. *Opt. Express* **26**, 27668–27682 (2018).

45. Zhang, W., Cui, X. & Martin, O. J. F. Local field enhancement of an infinite conical metal tip illuminated by a focused beam. *J. Raman Spectrosc.* **40**, 1338–1342 (2009).

46. Sanders, A. *et al.* Understanding the plasmonics of nanostructured atomic force microscopy tips. *Appl. Phys. Lett.* **109**, 153110 (2016).

47. Berweger, S., Atkin, J. M., Olmon, R. L. & Raschke, M. B. Light on the Tip of a Needle: Plasmonic Nanofocusing for Spectroscopy on the Nanoscale. *J. Phys. Chem. Lett.* **3**, 945–952 (2012).

48. Giugni, A. *et al.* Adiabatic nanofocusing: spectroscopy, transport and imaging investigation of the nano world. *J. Opt.* **16**, 114003 (2014).

49. Schmucker, S. W. *et al.* Field-directed sputter sharpening for tailored probe materials and atomic-scale lithography. *Nat. Commun.* **3**, 935 (2012).

50. Böckmann, H. *et al.* Near-Field Manipulation in a Scanning Tunneling Microscope Junction with Plasmonic Fabry-Pérot Tips. *Nano Lett.* **19**, 3597–3602 (2019).

51. Zhuang, X., Miranda, P. B., Kim, D. & Shen, Y. R. Mapping molecular orientation and conformation at interfaces by surface nonlinear optics. *Phys. Rev. B* **59**, 12632–12640 (1999).

52. Dalstein, L., Revel, A., Humbert, C. & Busson, B. Nonlinear optical response of a gold surface in the visible range: A study by two-color sum-frequency generation spectroscopy. I. Experimental determination. *J. Chem. Phys.* **148**, 134701 (2018).

53. Fellows, A. P. *et al.* Obtaining extended insight into molecular systems by probing multiple

pathways in second-order nonlinear spectroscopy. *J. Chem. Phys.* **159**, 164201 (2023).

54. Dall'Agnol, F. F. & Mammana, V. P. Solution for the electric potential distribution produced by sphere-plane electrodes using the method of images. *Rev. Bras. Ensino Fís.* **31**, 3503.1-3503.9 (2009).

55. Wang, H.-F., Gan, W., Lu, R., Rao, Y. & Wu, B.-H. Quantitative spectral and orientational analysis in surface sum frequency generation vibrational spectroscopy (SFG-VS). *Int. Rev. Phys. Chem.* **24**, 191–256 (2005).